 \documentclass[final,authoryear,times,twocolumn]{elsarticle}

\usepackage{amssymb}

\biboptions{sort&compress}

\journal{Applied Radiation and Isotopes}

\begin{document}

\begin{frontmatter}



\title{Activation cross-sections of deuteron induced reactions on natGd up to 50 MeV}


\author[1]{F. T\'ark\'anyi}
\author[1]{S. Tak\'acs}
\author[1]{F. Ditr\'oi\corref{*}}
\author[1]{J. Csikai}
\author[2]{A. Hermanne}
\author[3]{A.V. Ignatyuk}
\cortext[*]{Corresponding author: ditroi@atomki.hu}

\address[1]{Institute for Nuclear Research of the Hungarian Academy of Sciences (ATOMKI),  Debrecen, Hungary}
\address[2]{Cyclotron Laboratory, Vrije Universiteit Brussel (VUB), Brussels, Belgium}
\address[3]{Institute of Physics and Power Engineering (IPPE), Obninsk, Russia}

\begin{abstract}
Activation cross-sections are presented for the first time for $^{nat}Gd$(d,xn)$^{161,160,156(m+),154,154m1,154m2,153,152(m+),151(m+)}Tb$, $^{nat}Gd$(d,x)$^{159,153,151}Gd$ and $^{nat}Gd$(d,x)$^{156}Eu$ reactions from their respective thresholds up to 50 MeV. The cross-sections were measured by the stacked-foil irradiation technique and by using high resolution  $\gamma$-ray spectrometry. The measured values were compared with the results of theoretical models calculated by the computer codes ALICE-D, EMPIRE-D and TALYS (data from TENDL library). Integral yields of the reaction products were deduced from the excitation functions.
\end{abstract}


\end{frontmatter}


\section{Introduction}
\label{1}
In a previous paper \citep{TF2013} we reported cumulative cross-sections for the formation of medically used $^{161}Tb$ radioisotope in the bombardment of $^{nat}Gd$ with deuterons up to 50 MeV. In the frame of our ongoing systematic study of deuteron induced reactions for different applications, cross-sections of other radionuclides were also determined during that experiment. Terbium offers four clinically interesting radioisotopes with complementary physical decay characteristics: $^{149}Tb$, $^{152}Tb$, $^{155}Tb$, and $^{161}Tb$ \citep{Muller}. $^{153}Gd$ also has extensive use in nuclear medicine. Comparing the results of theoretical model codes with the experimental results, it turned out that the description of the (d,p) reaction is still problematic for the theory. While the importance of the deuteron induced reaction is rising, the corresponding experimental database is poor compared to that of protons. Taking into account that no earlier data were reported on activation cross-sections of deuteron induced reactions on Gd, we thought these results have or will have value for applications and for development of nuclear reaction codes. In this paper these excitation functions are presented. 

\section{Experiment and data evaluation}
\label{2}
Details of the experimental, as well as the data-analysis procedures, are described in our above mentioned previous work \citep{TF2013}. For the sake of completeness, the main experimental parameters and methods used for the two performed experiments on Gd are given here (Table 1) \citep{TF2001}, while the main parameters and data evaluation methods are collected in Table 2 (\citep{Andersen, Bonardi,  Canberra, Error, Kinsey, Pritychenko, Szekely, TF1991}. The used nuclear data of the produced radioisotopes are presented in Table 3 \citep{NuDat, Pritychenko}. 
The figure of the re-measured excitation function of the $^{27}Al$(d,x)$^{24}Na$ reaction for the high energy irradiation  in comparison with the IAEA recommended data \citep{TF2001} can be found in our earlier publication on Sc, irradiated in the same experiment \citep{Hermanne2012}. The figure of the re-measured excitation function of the $^{nat}Ti$(d,x)$^{48}V$ is not presented here, but the shape and accuracy was the same as published before \citep{TF2007}.

\begin{table*}[t]
\tiny
\caption{Main experimental parameters}
\centering
\begin{center}
\begin{tabular}{|p{1.3in}|p{1.3in}|p{1.4in}|} \hline 
Incident particle & Deuteron & Deuteron  \\ \hline 
Method  & Stacked foil & Stacked foil \\ \hline 
Target composition & ${}^{nat}$Gd (83.9 $\mu$m)\newline  Al \textbf{(}98 $\mu$m)\newline Ti (11 $\mu$m)\newline \newline  & ${}^{nat}$Gd (83.9 $\mu$m)\newline Sc (105 $\mu$m)\newline Al (27 $\mu$m) \\ \hline 
Number of Gd targetfoils & 9 & 20 \\ \hline 
Accelerator & CGR 560 cyclotron of Vrije Universiteit Brussels & Cyclone 90 cyclotron of the UniversitéCatholique in Louvain la Neuve (LLN)  \\ \hline 
Nominal energy & 21 MeV & 50 MeV \\ \hline 
Irradiation time & 110 min & 42 min \\ \hline 
Beam current (in Faraday) & 60 nA & 115 nA \\ \hline 
Monitor reaction, [recommended values]  & ${}^{nat}$Ti(d,x)${}^{48}$V reaction \citep{TF2001}\newline (re-measured over the whole energy range) & ${}^{27}$Al(d,x)${}^{24}$Na  reaction \citep{TF2001}\newline (re-measured over the whole energy range) \\ \hline 
Monitor target and thickness & ${}^{nat}$Ti, 11 $\mu$m & ${}^{nat}$Al, 27 $\mu$m \\ \hline 
detector & HpGe & HpGe \\ \hline 
Chemical separation & no & no \\ \hline 
g-spectra measurements & 2 series & 3 series \\ \hline 
Cooling times & 1.5-4.6 h,\newline 260-269 h\newline  & 2.2-5.3 h,\newline 20- 27h, \newline 197 - 227 h \\ \hline 
\end{tabular}
\end{center}
\end{table*}

\begin{table*}[t]
\tiny
\caption{Main parameters and methods of the data evaluation (with references)}
\centering
\begin{center}
\begin{tabular}{|p{1.3in}|p{2.0in}|p{1.5in}|} \hline 
\textbf{Parameter} & \textbf{Method} & \textbf{Reference} \\ \hline 
Gamma spectra evaluation & Genie 2000, Forgamma & \citep{Canberra, Szekely} \\ \hline 
Determination of beam intensity & Faraday cup (preliminary)\newline Fitted monitor reaction (final) &  \citep{TF1991} \\ \hline 
Decay data (see Table 3) & NUDAT 2.6  & \citep{Kinsey}  \\ \hline 
Reaction Q-values(see Table 3) & Q-value calculator & (Pritychenko and Sonzogni, 2003) \\ \hline 
Determination of beam energy & Andersen (preliminary)\newline Fitted monitor reaction (final) & \citep{Andersen} \newline \citep{TF1991} \\ \hline 
Uncertainty of energy & Cumulative effects of possible uncertainties\newline (nominal energy, target thickness, energy straggling,  correction to monitor reaction) &  \\ \hline 
Cross-sections & Isotopic and elemental cross-sections &  \\ \hline 
Uncertainty of cross-sections & sum in quadrature of all individual contributions\newline beam-current (7\%)\newline beam-loss corrections  (max.  1.5\%)\newline target thickness (1\%)\newline detector efficiency (5-7\%)\newline photo peak area determination and\newline counting statistics (1-20 \%)  \newline  & (International-Bureau-of-Weights-and-Measures, 1993) \\ \hline 
Yield & Physical yield & \citep{Bonardi} \\ \hline 
\end{tabular}

\end{center}
\end{table*}

\begin{table*}[t]
\tiny
\caption{Decay characteristic and contributing reactions for production of ${}^{161,160,}$${}^{156}$${}^{(m+),}$${}^{154,154m1,154m2,153,152,151}$Tb, ${}^{159,153,151}$Gd, ${}^{56}$Eu}
\centering
\begin{center}
\begin{tabular}{|p{0.6in}|p{0.5in}|p{0.6in}|p{0.5in}|p{1.2in}|p{0.7in}|} \hline 
Nuclide\newline Decay mode & Half-life & E${}_{\gamma}$(keV) & I${}_{\gamma}$(\%) & Contributing reaction & Q-value\newline (keV) \\ \hline 
\textbf{${}^{1}$${}^{61}$Tb\newline }$\beta $${}^{-}$: 100 \%\textbf{} & 6.89 d & 25.65135\newline 48.91533\newline 57.1917\newline 74.56669 & 23.2\newline 17.0\newline 1.79\newline 10.2 & ${}^{1}$${}^{60}$Gd(d,n)\newline ${}^{160}$Gd(d,p)${}^{161}$Gd›${}^{161}$Tb & 4584.3\newline 3410.83 \\ \hline 
\textbf{${}^{1}$${}^{60}$Tb\newline }$\beta $${}^{-}$: 100 \%\textbf{} & 72.3 d & 86.7877\newline 197.0341\newline 215.6452\newline 298.57 83\newline 879.378\newline 962.311\newline 966.166\newline 1177.954 & 13.2\newline 5.18\newline 4.02\newline 26.1\newline 30.1\newline 9.81\newline 25.1\newline 14.9 & ${}^{1}$${}^{60}$Gd(d,2n) & -3112.31 \\ \hline 
\textbf{${}^{1}$${}^{56}$Tb\newline }$\varepsilon $: 100 \% & 5.35 d & 88.97\newline 199.19\newline 262.54\newline 296.49\newline 356.38\newline 422.34\newline 534.29\newline 1065.11\newline 1154.07\newline 1159.03\newline 1222.44\newline 1421.67 & 18.0\newline 41.0\newline 5.8\newline 4.5\newline 13.6\newline 8.0\newline 67\newline 10.8\newline 10.4\newline 7.2\newline 31\newline 12.2 & ${}^{1}$${}^{55}$Gd(d,n)\newline ${}^{1}$${}^{56}$Gd(d,2n)\newline ${}^{1}$${}^{57}$Gd(d,3n)\newline ${}^{1}$${}^{58}$Gd(d,4n)\newline ${}^{1}$${}^{60}$Gd(d,6n) & 3085.25\newline -5451.1\newline -11810.96\newline -19748.35\newline -33143.03 \\ \hline 
\textbf{${}^{1}$${}^{55}$Tb\newline }$\varepsilon $: 100 \% & 5.32 d & 86.55\newline 105.318\newline 148.64\newline 161.29\newline 163.28\newline 180.08\newline 340.67\newline 367.36 & 32.0\newline 25.1\newline 2.65\newline 2.76\newline 4.44\newline 7.5\newline 1.18\newline 1.48 & ${}^{1}$${}^{54}$Gd(d,n)\newline ${}^{1}$${}^{55}$Gd(d,2n)\newline ${}^{1}$${}^{56}$Gd(d,3n)\newline ${}^{1}$${}^{57}$Gd(d,4n)\newline ${}^{1}$${}^{58}$Gd(d,5n)\newline ${}^{1}$${}^{60}$Gd(d,7n) & 2608.53\newline -3826.7\newline -12363.0\newline -18722.9\newline -26660.3\newline -40055.0 \\ \hline 
\textbf{${}^{154m1}$Tb\newline }e${}^{+}$+b${}^{+}$ (78.2 \%)  \newline IT( 21.8 \%)\textbf{} & 9.4 h & 518.011\newline 540.18\newline 649.564\newline 873.190\newline 996.262\newline 1004.725 & 6.1\newline 20\newline 10.9\newline 9.2\newline 8.6\newline 10.9\newline  & ${}^{1}$${}^{54}$Gd(d,2n)\newline ${}^{1}$${}^{55}$Gd(d,3n)\newline ${}^{1}$${}^{56}$Gd(d,4n)\newline ${}^{1}$${}^{57}$Gd(d,5n)\newline ${}^{1}$${}^{58}$Gd(d,6n)\newline ${}^{1}$${}^{60}$Gd(d,8n) & -6556.6${}^{*}$\newline -12991.8${}^{*}$\newline -21528.1${}^{*}$\newline -27888.0${}^{*}$\newline -35825.4*\newline -49220.1${}^{*}$ \\ \hline 
\textbf{${}^{1}$${}^{54m2}$Tb\newline \newline }e${}^{+}$+b${}^{+}$( 98.2 \%) \newline IT (1.8 \%)\newline \textbf{} & 22.7 h & 123.071\newline 225.94\newline 346.643\newline 1004.725\newline 1419.81 & 43 \newline 26.8\newline 69\newline 7.1\newline 46 & ${}^{1}$${}^{54}$Gd(d,2n)\newline ${}^{1}$${}^{55}$Gd(d,3n)\newline ${}^{1}$${}^{56}$Gd(d,4n)\newline ${}^{1}$${}^{57}$Gd(d,5n)\newline ${}^{1}$${}^{58}$Gd(d,6n)\newline ${}^{1}$${}^{60}$Gd(d,8n) & -6556.6${}^{*}$\newline -12991.8${}^{*}$\newline -21528.1${}^{*}$\newline -27888.0${}^{*}$\newline -35825.4*\newline -49220.1${}^{*}$ \\ \hline 
\textbf{${}^{1}$${}^{54g}$Tb\newline }$\varepsilon $: 100 \%~\textbf{} & 21.5 h & 123.07\newline 247.94\newline 557.60\newline 692.41\newline 704.90\newline 722.12\newline ~873.21\newline 996.24\newline 1274.436\newline 1291.31 & 26\newline 1.7\newline 5.4\newline 3.18\newline 4.8\newline 7.7\newline 5.3\newline 4.9\newline 10.5\newline 6.9 & ${}^{1}$${}^{54}$Gd(d,2n)\newline ${}^{1}$${}^{55}$Gd(d,3n)\newline ${}^{1}$${}^{56}$Gd(d,4n)\newline ${}^{1}$${}^{57}$Gd(d,5n)\newline ${}^{1}$${}^{58}$Gd(d,6n)\newline ${}^{1}$${}^{60}$Gd(d,8n) & -6556.6\newline -12991.8\newline -21528.1\newline -27888.0\newline -35825.4\newline -49220.1 \\ \hline 
\textbf{${}^{1}$${}^{53}$Tb\newline }$\varepsilon $: 100 \%\textbf{\newline } & 2.34 d & 102.255\newline 109.758\newline 170.42\newline 212.00\newline 249.55 & 6.4\newline 6.8\newline 6.3\newline 31.0\newline 2.33 & ${}^{1}$${}^{52}$Gd(d,n)\newline ${}^{1}$${}^{54}$Gd(d,3n)\newline ${}^{1}$${}^{55}$Gd(d,4n)\newline ${}^{1}$${}^{56}$Gd(d,5n)\newline ${}^{1}$${}^{57}$Gd(d,6n)\newline ${}^{1}$${}^{58}$Gd(d,7n) & 1671.05\newline -13470.62\newline -19905.86\newline -28442.21\newline -34802.07\newline -42739.46 \\ \hline 
\textbf{${}^{1}$${}^{52}$Tb\newline }$\varepsilon $: 100 \%\textbf{\newline } & 17.5 h & 271.08\newline 344.28\newline 411.08\newline 586.29\newline 764.88\newline 778.86\newline 974.1\newline 1109.2\newline 1299.11\underbar{} & 8.6\newline 65\newline 4.1\newline 9.4\newline 2.9\newline 5.8\newline 3.1\newline 2.7\newline 2.15 & ${}^{1}$${}^{52}$Gd(d,2n)\newline ${}^{1}$${}^{54}$Gd(d,4n)\newline ${}^{1}$${}^{55}$Gd(d,5n)\newline ${}^{1}$${}^{56}$Gd(d,6n)\newline ${}^{1}$${}^{57}$Gd(d,7n) & -6996.9\newline -22138.6\newline -28573.8\newline -37110.2\newline -43470.0 \\ \hline 
\textbf{${}^{151}$Tb\newline }$\alpha $: 0.0095\newline (0.0095 \%)\newline $\varepsilon $: 99.9905 \%\textbf{} & 17.609 h & 108.088\newline 180.186\newline 251.863\newline 287.357\newline 395.444\newline 443.879\newline 479.357\newline 587.46\newline 616.561\newline 731.227 & 24.3\newline 11.5\newline 26.3\newline 28.3\newline 10.8\newline 10.8\newline 15.4\newline 15.6\newline 10.4\newline 7.7 & ${}^{1}$${}^{52}$Gd(d,3n)\newline ${}^{1}$${}^{54}$Gd(d,5n)\newline ${}^{1}$${}^{55}$Gd(d,6n)\newline ${}^{1}$${}^{56}$Gd(d,7n)\newline ${}^{1}$${}^{57}$Gd(d,8n) & -14161.58\newline -29303.27\newline -35738.5\newline -44274.85\newline -50634.71 \\ \hline 
\end{tabular}
\end{center}
\end{table*}

\setcounter{table}{2}
\begin{table*}[t]
\tiny
\caption{Table 3 cont.}
\label{Table 3 cont.}
\centering
\begin{center}
\begin{tabular}{|p{0.6in}|p{0.5in}|p{0.6in}|p{0.5in}|p{1.2in}|p{0.7in}|} \hline 
Nuclide\newline Decay mode & Half-life & E${}_{\gamma}$(keV) & I${}_{\gamma}$(\%) & Contributing reaction & Q-value\newline (keV) \\ \hline 

\textbf{${}^{1}$${}^{59}$Gd\newline }$\beta $${}^{-}$: 100 \%\textbf{} & 18.479 h & 363.5430 & 11.78 & ${}^{1}$${}^{58}$Gd(d,p)\newline ${}^{1}$${}^{60}$Gd(d,p2n)\newline ${}^{159}$Eu decay & 3718.644\newline -9676.04 \\ \hline 
\textbf{${}^{1}$${}^{53}$Gd\newline }$\varepsilon $: 100 \%\textbf{} & 240.4 d & 97.43100\newline 103.18012 & 29.0\newline 21.1 & ${}^{1}$${}^{52}$Gd(d,p)\newline ${}^{1}$${}^{54}$Gd(d,p2n)\newline ${}^{1}$${}^{55}$Gd(d,p3n)\newline ${}^{1}$${}^{56}$Gd(d,p4n)\newline ${}^{1}$${}^{57}$Gd(d,p5n)\newline ${}^{1}$${}^{58}$Gd(d,p6n) & 4022.394\newline -11119.3\newline -17554.53\newline -26090.88\newline -32450.74\newline -40388.13 \\ \hline 
\textbf{${}^{1}$${}^{51}$Gd\newline }$\varepsilon $: 100 \%\newline $\alpha $: 0.8E-6 \%\textbf{} & 123.9 d & 153.60\newline 174.70\newline 243.29 & 6.2\newline 2.96\newline 5.6 & ${}^{1}$${}^{52}$Gd(d,p2n)\newline ${}^{1}$${}^{54}$Gd(d,p4n)\newline ${}^{1}$${}^{55}$Gd(d,p5n)\newline ${}^{1}$${}^{56}$Gd(d,p6n)\newline ${}^{1}$${}^{57}$Gd(d,p7n)\newline ${}^{1}$${}^{58}$Gd(d,p8n)\newline ${}^{1}$${}^{60}$Gd(d,p10n) & -10814.23\newline -25955.91\newline -32391.15\newline -40927.5\newline -47287.36\newline -55224.75\newline X \\ \hline 
\textbf{${}^{1}$${}^{56}$Eu\newline }$\beta $${}^{-}$: 100 \%\textbf{} & 15.19 d & 646.29\newline 723.47\newline 811.77\newline 1079.16\newline 1153.67\newline 1154.08\newline 1230.71\newline 1242.42 & 6.3\newline 5.4\newline 9.7\newline 4.6\newline 6.8\newline 4.7\newline 8.0\newline 6.6 & ${}^{156}$Gd(d,2p)\newline ${}^{157}$Gd(d,2pn)\newline ${}^{158}$Gd(d,2p2n)\newline ${}^{1}$${}^{60}$Gd(d,2p4n) & -3891.35\newline -10251.21\newline -18188.6\newline -31583.28 \\ \hline 

\end{tabular}
\end{center}

\begin{flushleft}
\footnotesize{\noindent The Q-values refer to formation of the ground state and are obtained fromthe Q-value calculation tool of Pritychenko et al. \citep{Pritychenko}. When complex particles are emitted instead of individual protons and neutrons the Q-values have to be decreased by the respective binding energies of the compound particles: np-d, +2.2 MeV; 2np-t, +8.48 MeV; n2p-${}^{3}$He, +7.72 MeV; 2n2p-$\alpha$, +28.30 MeV. 

\noindent *In case of ${}^{154m1}$Tb and ${}^{154m2}$Tb  the Table 3 includes   Q values of the ground state, because level energies of the isomeric states are not known.}
\end{flushleft}
\end{table*}

\section{Model calculations}
\label{3}
For theoretical estimation of the excitation functions involved in this experiment the updated ALICE-IPPE-D \citep{Dityuk}  and EMPIRE-D \citep{Herman} codes were used. In the modified  versions of the codes a simulation of direct (d,p) and (d,t) transitions by the general relations for a nucleon transfer probability in the continuum is included through an energy dependent enhancement factor for the corresponding transitions based on systematics of experimental data \citep{Ignatyuk}. Since ALICE-IPPE cannot provide direct cross-section results for excited isomeric states, the cross-section for any isomeric state was obtained by applying the isomeric ratios derived from the EMPIRE calculation to the total cross-sections calculated by ALICE-IPPE.
The experimental cross-section data were also compared with the results of the modified TALYS code \citep{Koning} taken from the TENDL-2011 and TENDL-2012 data libraries \citep{KoningTALYS}.

\section{Results and discussion}
\label{4}

\subsection{Cross-sections}
\label{4.1}
The cross-sections for radionuclides produced in deuteron bombardment of $^{nat}Gd$ target are tabulated in Tables 4 and 5. The experimental cross-section data are compared with the theory graphically in Figs 1-14. The cross-sections given for 160Tb and $^{161}Tb$ are considered to be isotopic cross-sections as the radionuclides can be induced mainly on $^{160}Gd$, as the contribution of (d,$\gamma$) reaction on $^{158}Gd$ is negligible. In all other cases so called "elemental cross-sections" were deduced on targets with natural isotopic composition. The values for $^{160}Tb$, $^{155}Tb$, $^{154m1}Tb$, $^{154m2}Tb$, $^{154g}Tb$, $^{153}Tb$ and in practice also for $^{156}Tb$ and $^{156}Eu$ represent production cross-sections via direct nuclear reactions only. In the cases, where significantly shorter-lived precursors (isomeric states (m+) and/or parent isotopes (cum)) are also produced, the cross-sections for the ground state were derived from activity measurements performed after the nearly complete decay of the precursor. These cross-sections are hence the sum for the direct production and formation via the decay of the precursors. The values for $^{156}Tb$, $^{152}Tb$ and $^{151}Tb$ contain contribution from the short-lived isomeric states (m+). Cross-sections are cumulative (cum), with contribution of the decay of parent radionuclides, in case of $^{161}Tb$, $^{159}Gd$, $^{153}Gd$ and $^{151}Gd$.
For the sake of completeness we include the results for the earlier published investigations for production of $^{161}Tb$ and $^{160}Tb$ both in graphical and in numerical form \citep{TF2013}.

\subsubsection{Activation cross-sections for production of $^{161}Tb$ (cum) ($T_{1/2}$ = 6.89 d)}
\label{4.1.1}
We are reproducing the earlier published experimental results (Fig. 1) for production of $^{161}Tb$(cum) (direct, through $^{160}Gd$(d,n) and decay of $^{161}Gd$ produced by $^{160}Gd$(d,p)$^{161}Gd$) to illustrate the agreement of the theoretical and experimental data for further discussion. 
The agreement is acceptable for ALICE-D and EMPIRE-D. The TALYS results are still too low, although the 2012 values are 10\% higher than 2011 values. More detailed discussion on the contributing reactions and on the application of $^{161}Tb$ can be found in the previous work \citep{TF2013}.

\begin{figure}[h]
\includegraphics[scale=0.3]{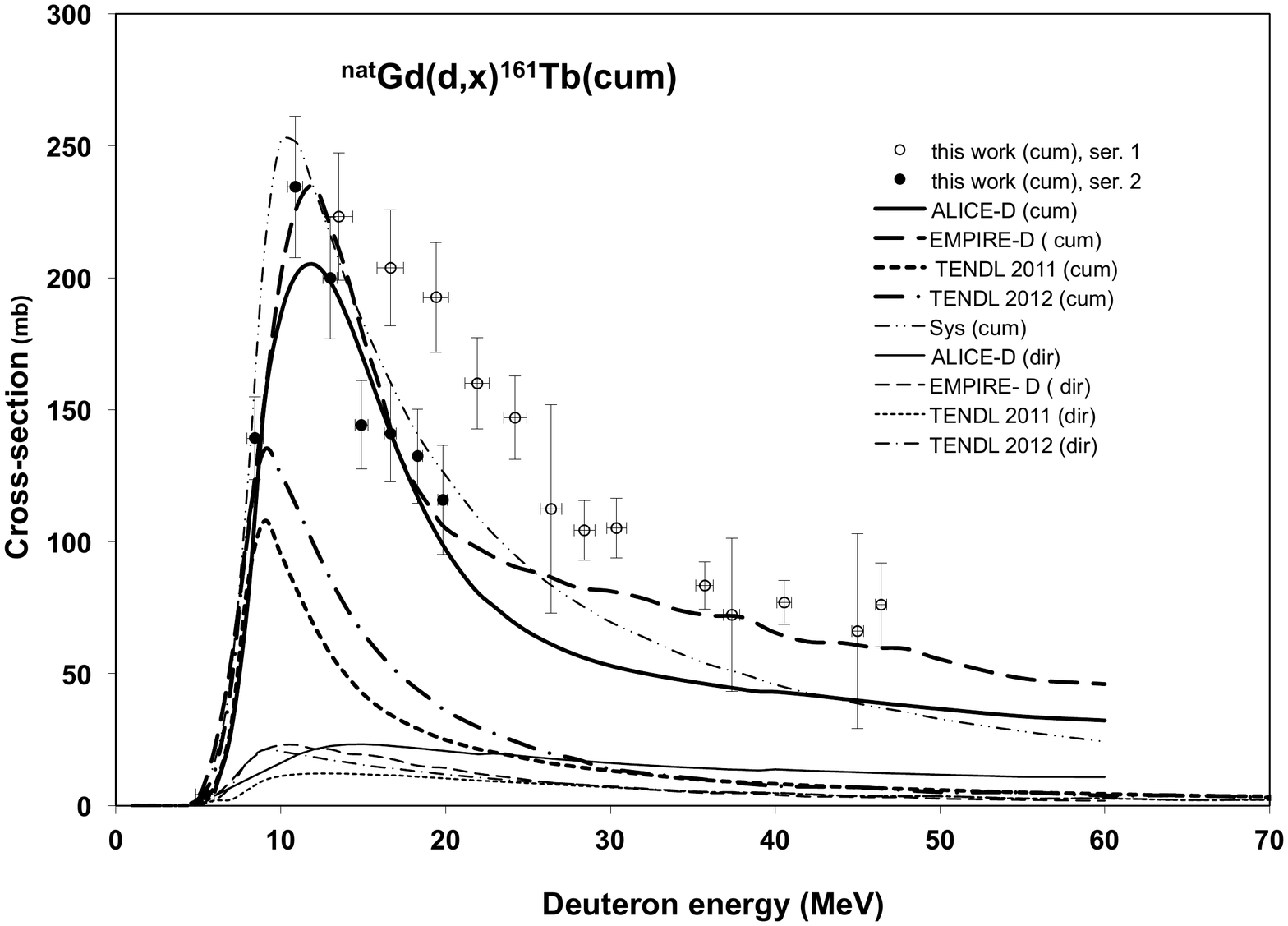}
\caption{Experimental and theoretical excitation functions of the $^{nat}Gd$(d,x)$^{161}Tb$ reaction }
\end{figure}

\subsubsection{Activation cross-sections for production of $^{160}Tb$ ($T_{1/2}$ = 72.3 d)}
\label{4.1.2}
The $^{160}Tb$ is produced directly through the $^{160}Gd$(d,2n) reaction. Fig. 2 illustrates the prediction capability of the theoretical codes including our ALICE-D and EMPIRE-D results. There are significant differences between the not adjusted theoretical results. There is no difference between TENDL-2011 and TENDL-2012. More detailed discussion was reported in \citep{TF2013}.

\begin{figure}[h]
\includegraphics[scale=0.3]{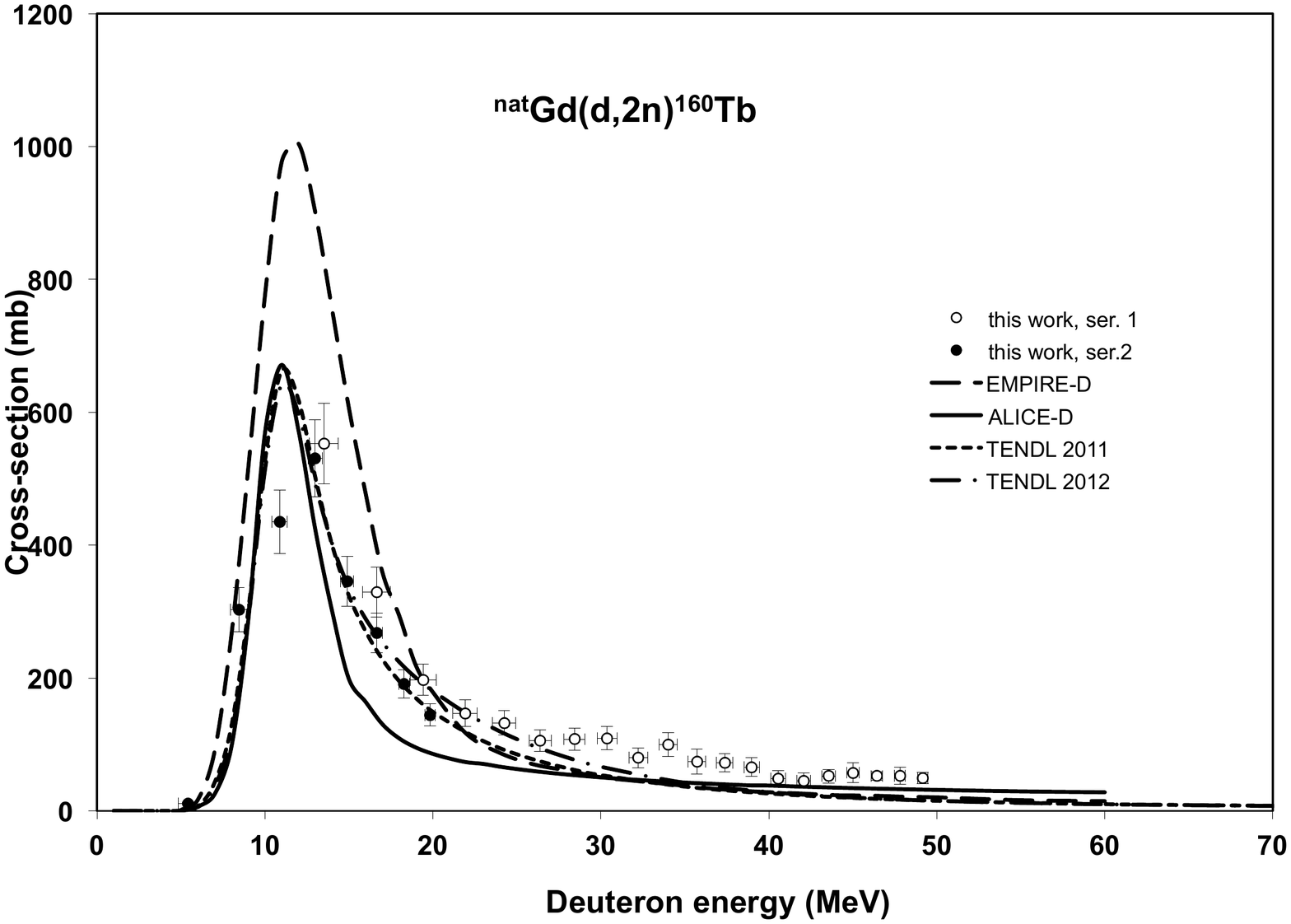}
\caption{Experimental and theoretical excitation function of the $^{nat}Gd$(d,2n)$^{160}Tb$ reaction}
\end{figure}

\subsubsection{Activation cross-sections for production of $^{156}Tb$(tot) ($T_{1/2}$=5.35 d) }
\label{4.1.3}
We have measured the production cross-section of the ground state of the $^{156}Tb$ (Fig. 3) after the decay of the isomeric states ($^{156m1}Tb$, IT: 100 \%, $T_{1/2}$ = 3 h and $^{156m2}Tb$, IT: 100 \%, $T_{1/2}$ = 24.4 h). All theoretical results follow the shape of the experimental excitation function. The EMPIRE-D results are significantly overestimating the experimental values (similar to the previous reaction).

\begin{figure}[h]
\includegraphics[scale=0.3]{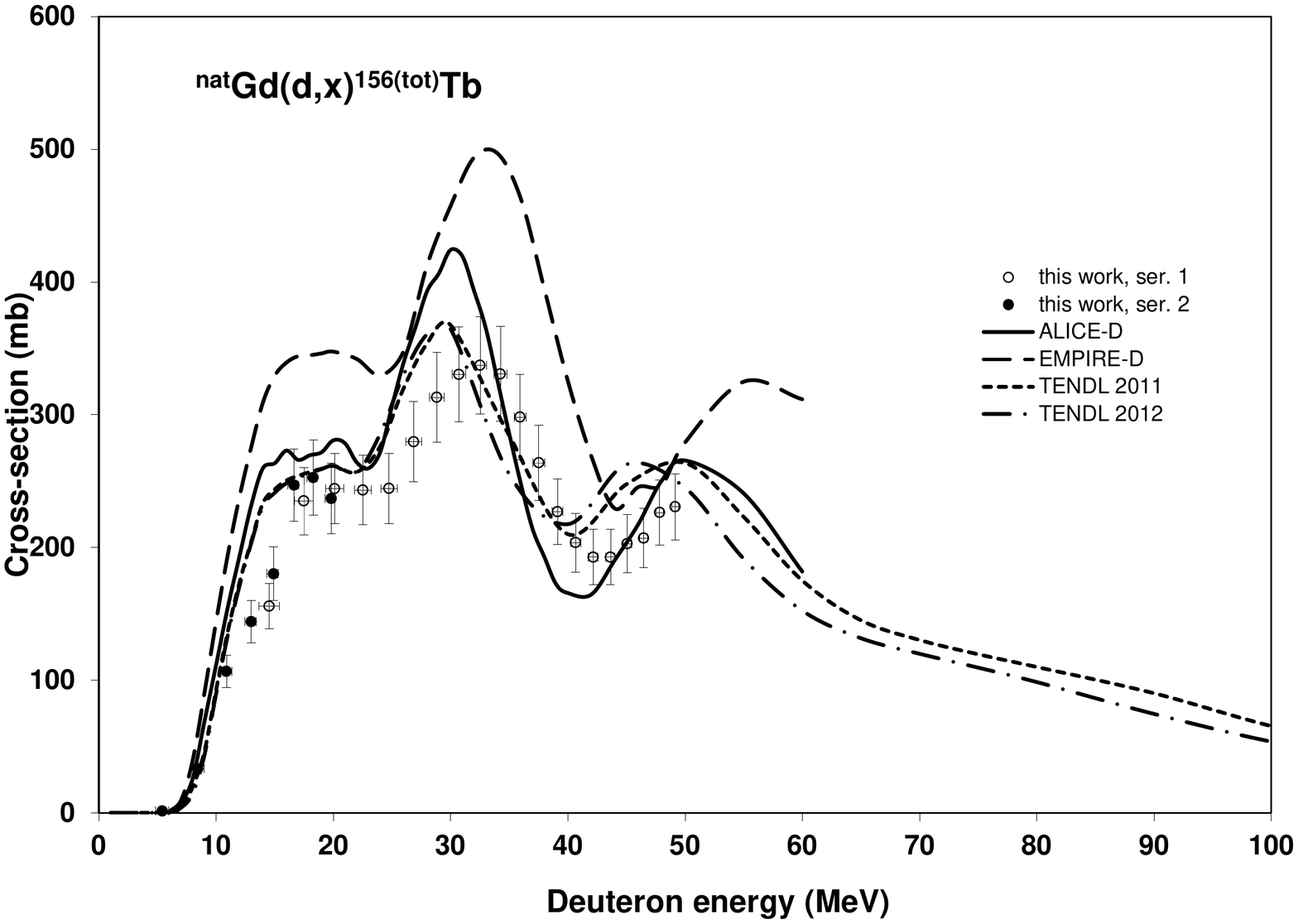}
\caption{Experimental and theoretical excitation function of the $^{nat}Gd$(d,x)$^{156}Tb$(m+) reaction}
\end{figure}

\subsubsection{Activation cross-sections for production of $^{155}Tb$ ($T_{1/2}$=5.32 d)}
\label{4.1.4}
According to Fig. 4 the ALICE-D and EMPIRE-D significantly overestimate the experimental data, while the TENDL results (nearly identical for 2011 and 2012) are in good accordance.

\begin{figure}[h]
\includegraphics[scale=0.3]{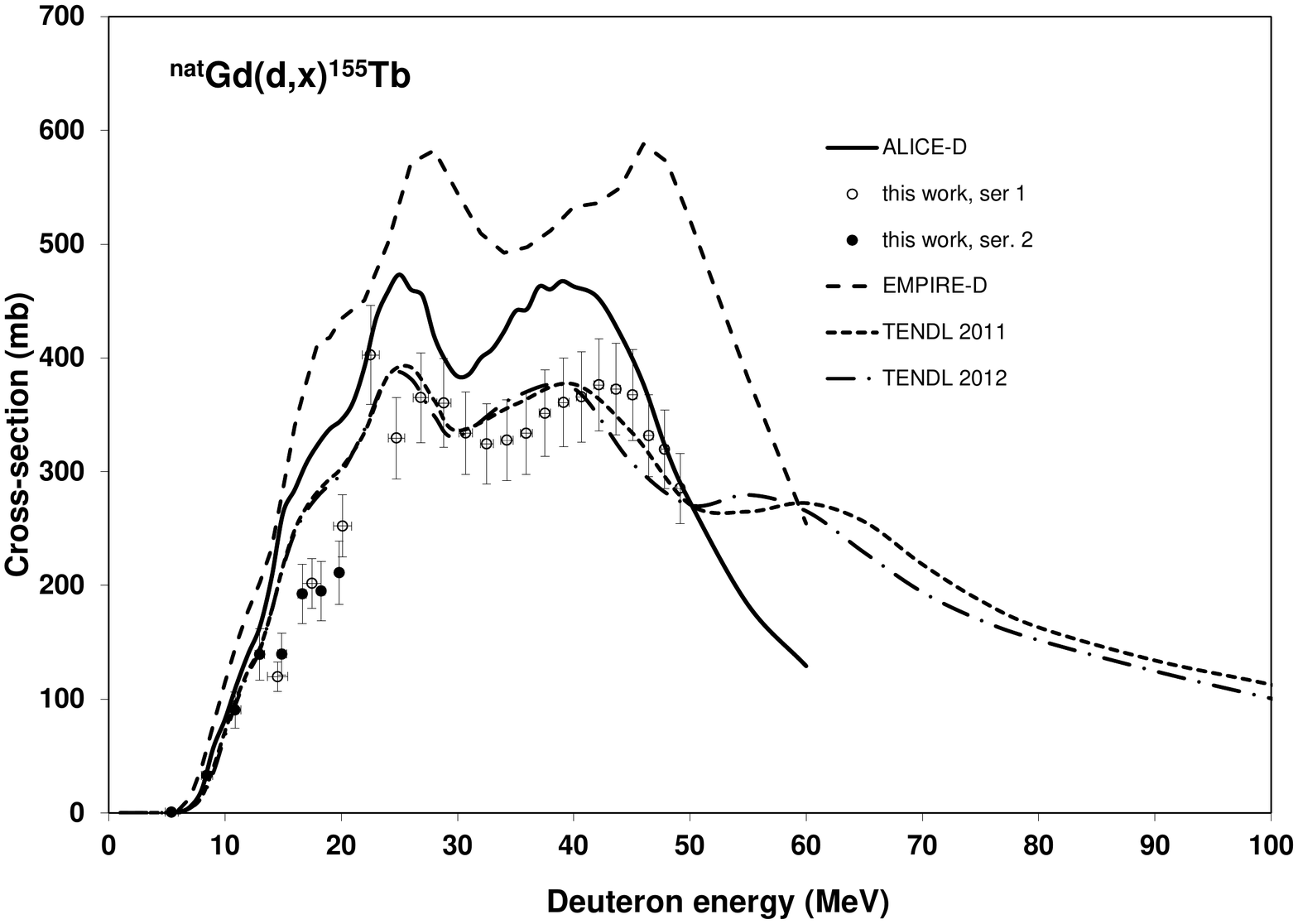}
\caption{Experimental and theoretical excitation function of the $^{nat}Gd$(d,x)$^{155}Tb$ reaction}
\end{figure}

\subsubsection{Activation cross-sections for production of $^{154m2}Tb$ ($T_{1/2}$=22.7 h)}
\label{4.1.5}
The $^{154}Tb$ has three, long-lived isomeric states. The cross-sections for direct production of the higher lying isomer ($T_{1/2}$ = 22.7 h, $J^{\pi} = 7^-$,  $\varepsilon^+ + \beta^+$ 98.2 \%, IT 1.8 \% to the $^{154m1}Tb$ (Ekström and Firestone, 1999)) are shown in Fig. 5 in comparison with the theoretical result of EMPIRE-D and ALICE-D. No data are available in the TENDL libraries for production of this isomeric state.

\begin{figure}[h]
\includegraphics[scale=0.3]{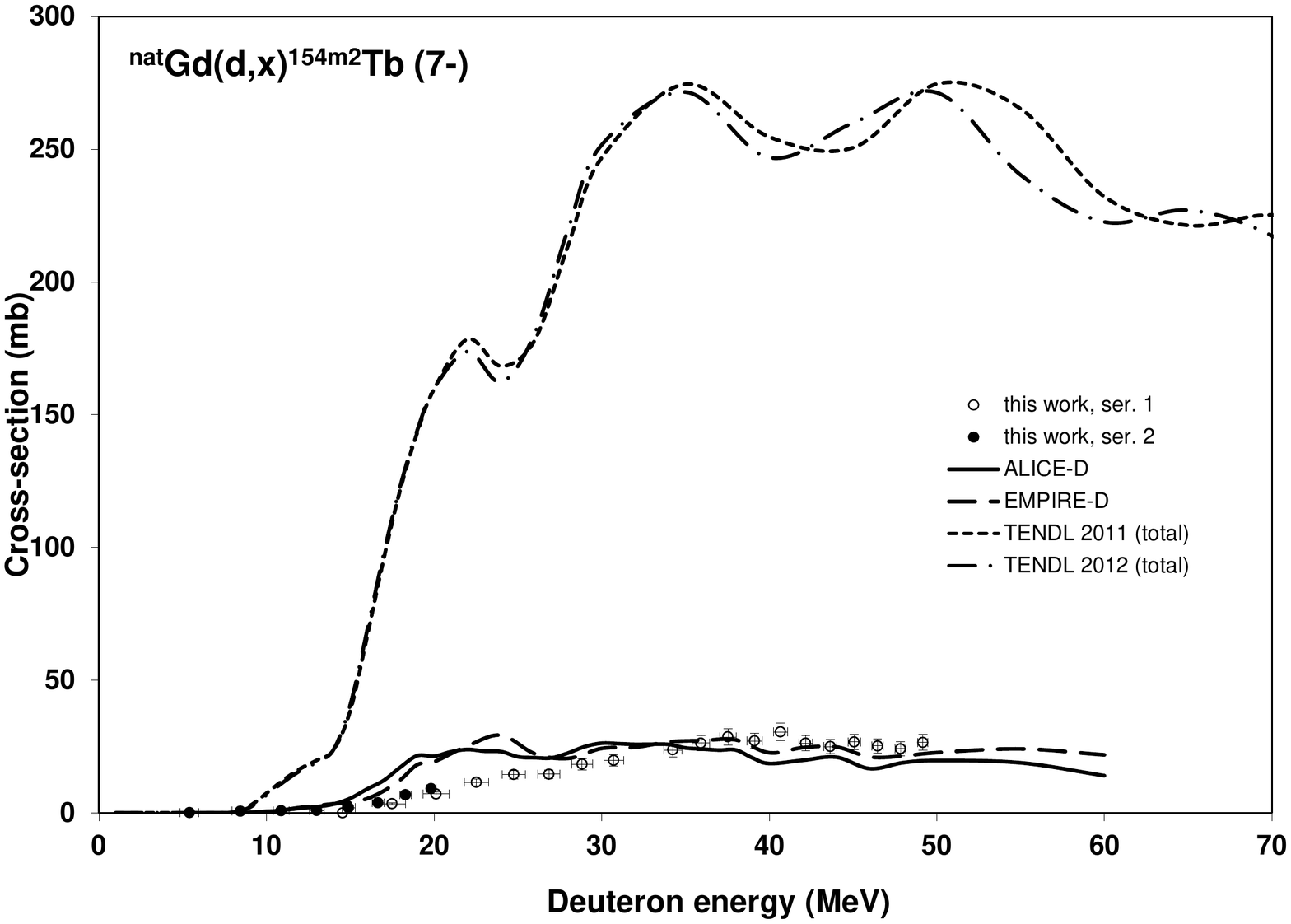}
\caption{Experimental and theoretical excitation function of the $^{nat}Gd$(d,x)$^{154m2}Tb$ reaction}
\end{figure}

\subsubsection{Activation cross-sections for production of $^{154m1}Tb$ ($T_{1/2}$=9.4 h)}
\label{4.1.6}
The cross-section of the second (lower-lying) isomeric state (9.4 h, $J^{\pi} = 3^-$,  $\varepsilon^+ + \beta^+$ +78.2 \%, IT 21.8 \% to $^{154g}Tb$ \citep{Ekstrom}) is shown in Fig. 6. It should be noted that there are new data for the half-life (9.994 h \citep{Rastrepina}, but it is not validated, therefore we used $T_{1/2}$= 9.4 h.
The $^{154m1}Tb$ is produced both directly and from the decay of $^{154m2}Tb$. The contribution from the decay of $^{154m2}Tb$ was corrected. The correction was small taking into account the low cross-sections and the low amount of the internal transition. The excitation function is shown in Fig 6 in comparison with the theory. The experimental data are significantly higher than the theoretical results. During our EMPIRE-D calculation some problems arose with the level scheme of $^{154}Tb$. There are no experimental data for the low-lying levels around the isomeric states. As such, the scheme used in calculations is artificial and based on the systematics of levels of the neighboring nuclei.

\begin{figure}[h]
\includegraphics[scale=0.3]{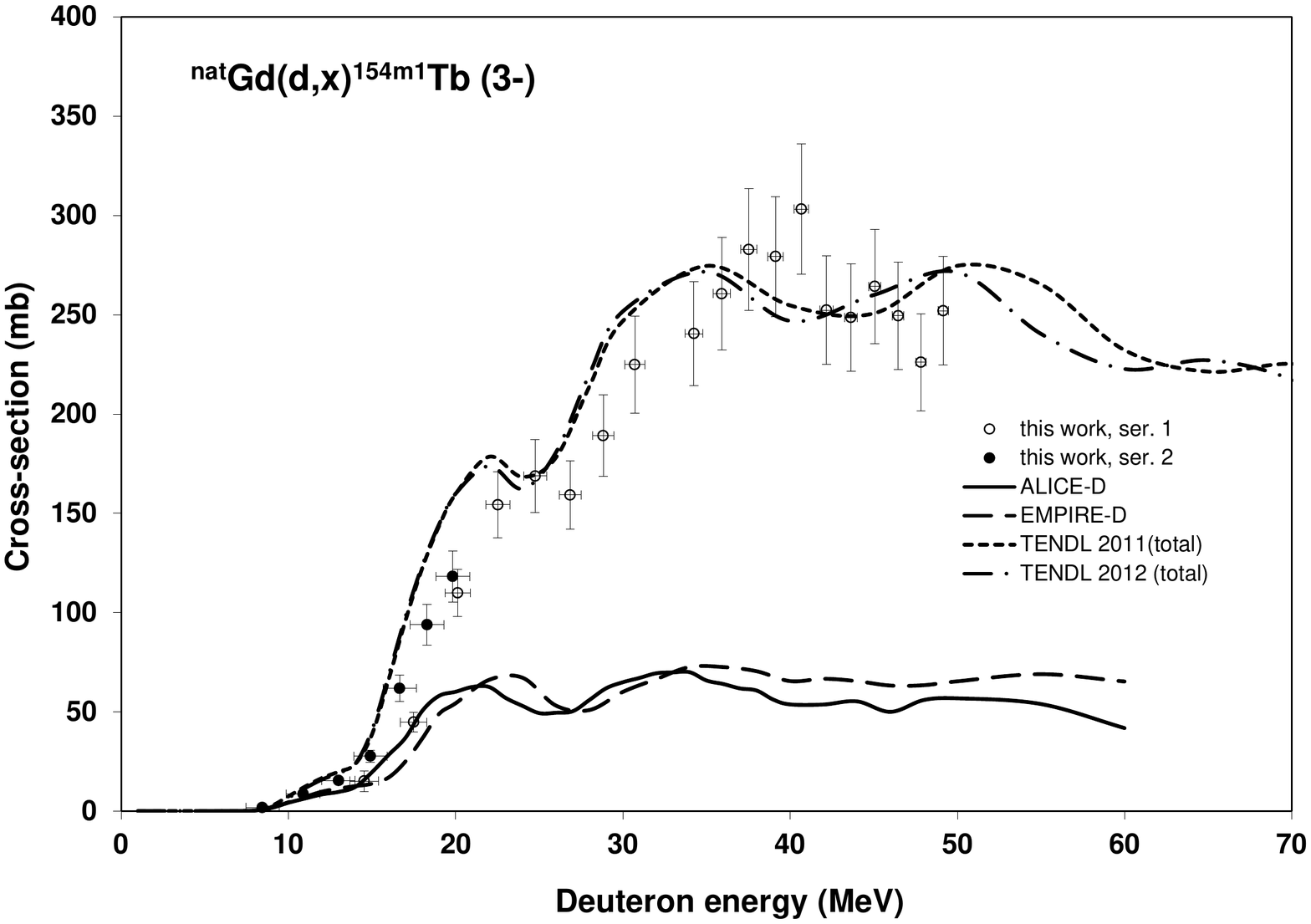}
\caption{Experimental and theoretical excitation function of the $^{nat}Gd$(d,x)$^{154m1}Tb$ reaction}
\end{figure}

\subsubsection{Activation cross-sections for production of $^{154g}Tb$(m1+) ($T_{1/2}$=21.5 h ) }
\label{4.1.7}
By using the late spectra we can deduce cumulative cross-sections for production of the ground state after the complete decay of the $^{154m1}Tb$ (Fig. 7). The contribution of the decay of $^{154m2}Tb$ to the production of the ground state is negligibly small compared to the direct production and to the contribution from the first isomeric state. The experimental data are systematically higher than the results of the EMPIRE-D. No data exist in the TENDL libraries for the isomeric states, only for the total production. 

\begin{figure}[h]
\includegraphics[scale=0.3]{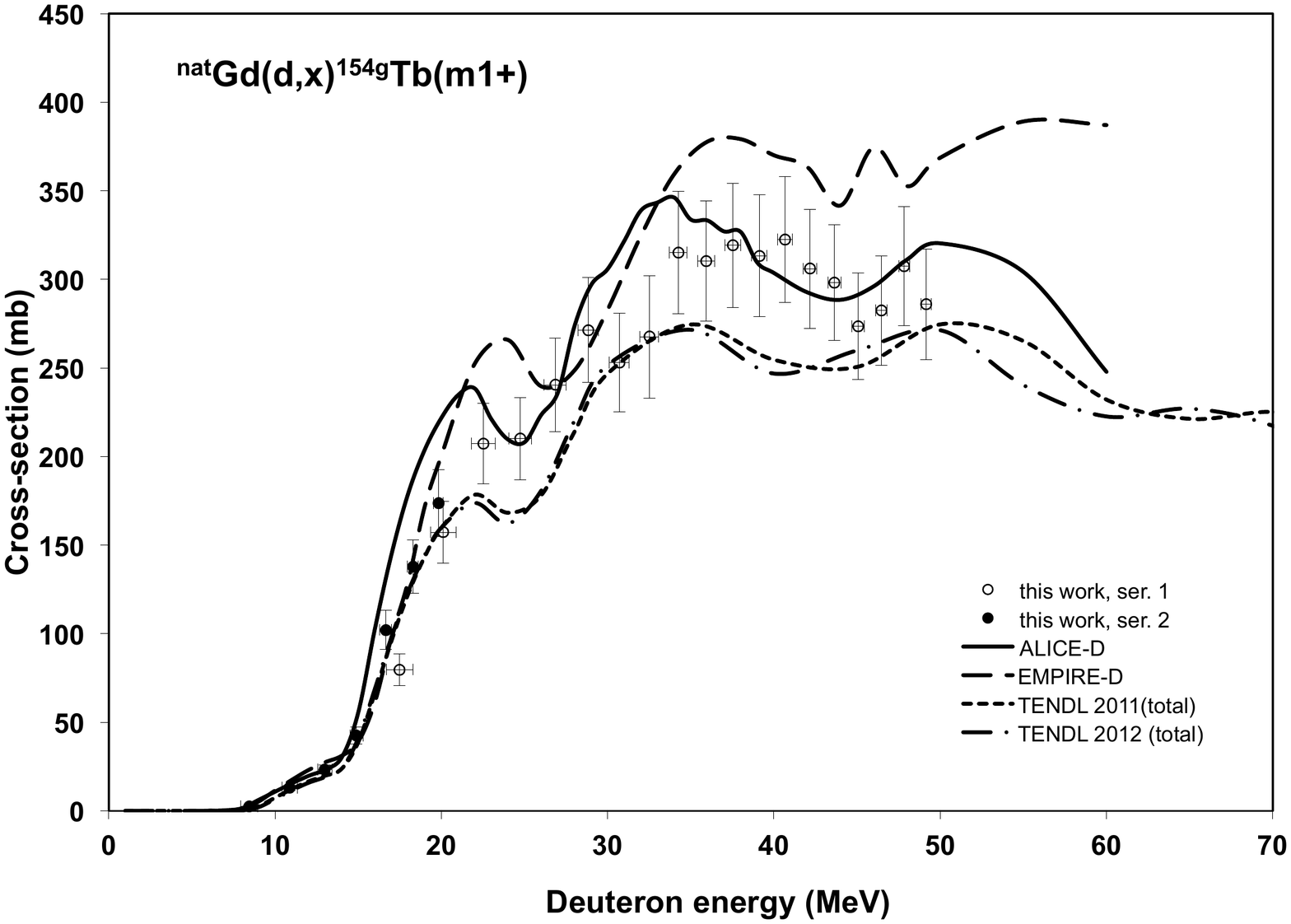}
\caption{Experimental and theoretical excitation function of the $^{nat}Gd$(d,x)$^{154g}Tb$ (m1+) reaction}
\end{figure}

\subsubsection{Activation cross-sections for production of $^{153}Tb$ ($T_{1/2}$=2.34 d)}
\label{4.1.8}
The agreement with the TENDL results is good (Fig. 8). There are large differences in the absolute values of the predictions by ALICE-D and EMPIRE-D at energies above 30 MeV. 

\begin{figure}[h]
\includegraphics[scale=0.3]{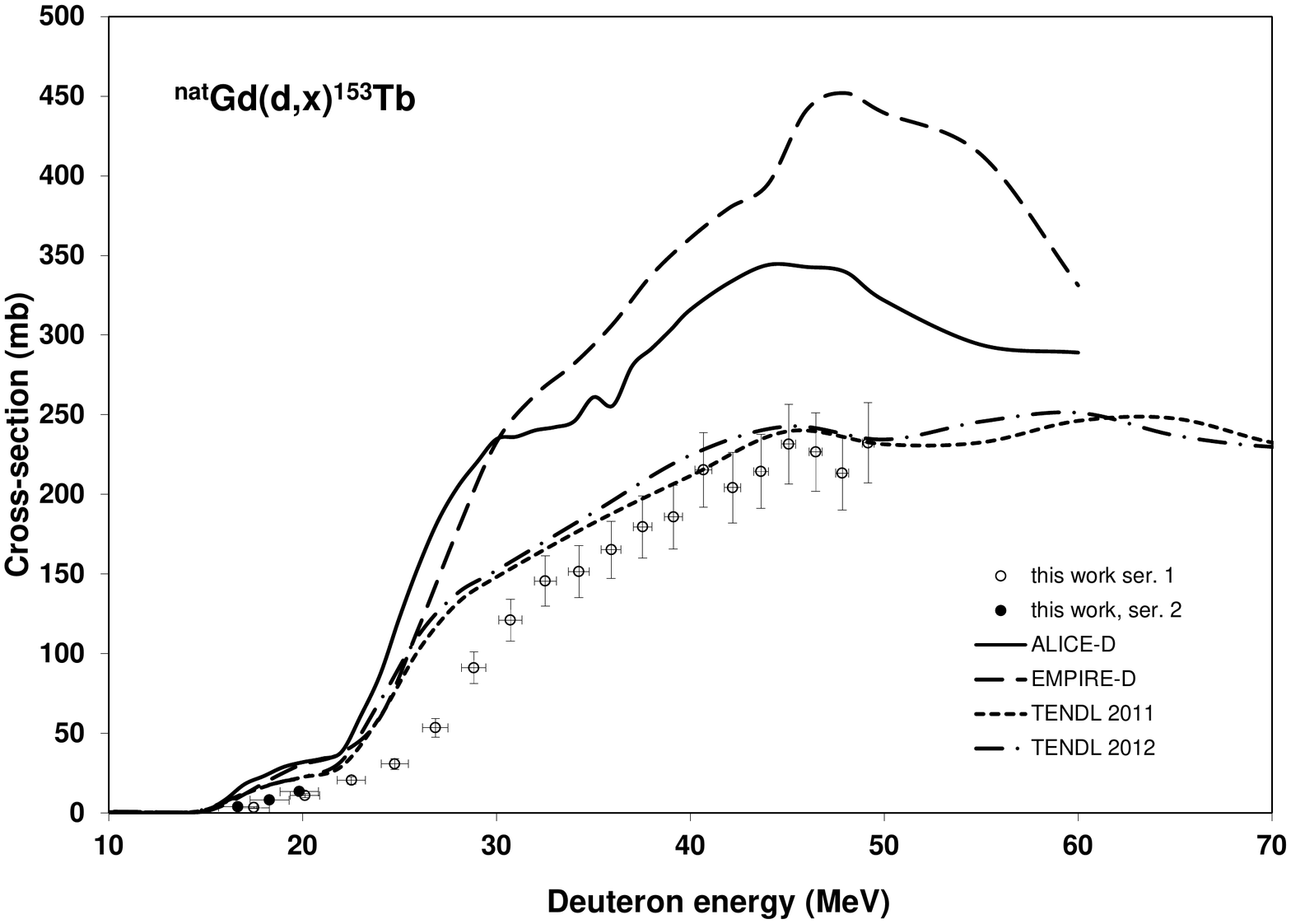}
\caption{Experimental and theoretical excitation function of the $^{nat}Gd$(d,x)$^{153}Tb$ reaction}
\end{figure}

\subsubsection{Activation cross-sections for production of $^{152}Tb$ (m+)( $T_{1/2}$=17.5 h)}
\label{4.1.9}
The excitation function was measured after the decay of the short-lived isomeric state (4.2 min, IT: 78.8 \%). The agreement with the TENDL library results (Fig. 9) is good, while ALICE-D and EMPIRE-D strongly overestimate again above 40 MeV. 

\begin{figure}[h]
\includegraphics[scale=0.3]{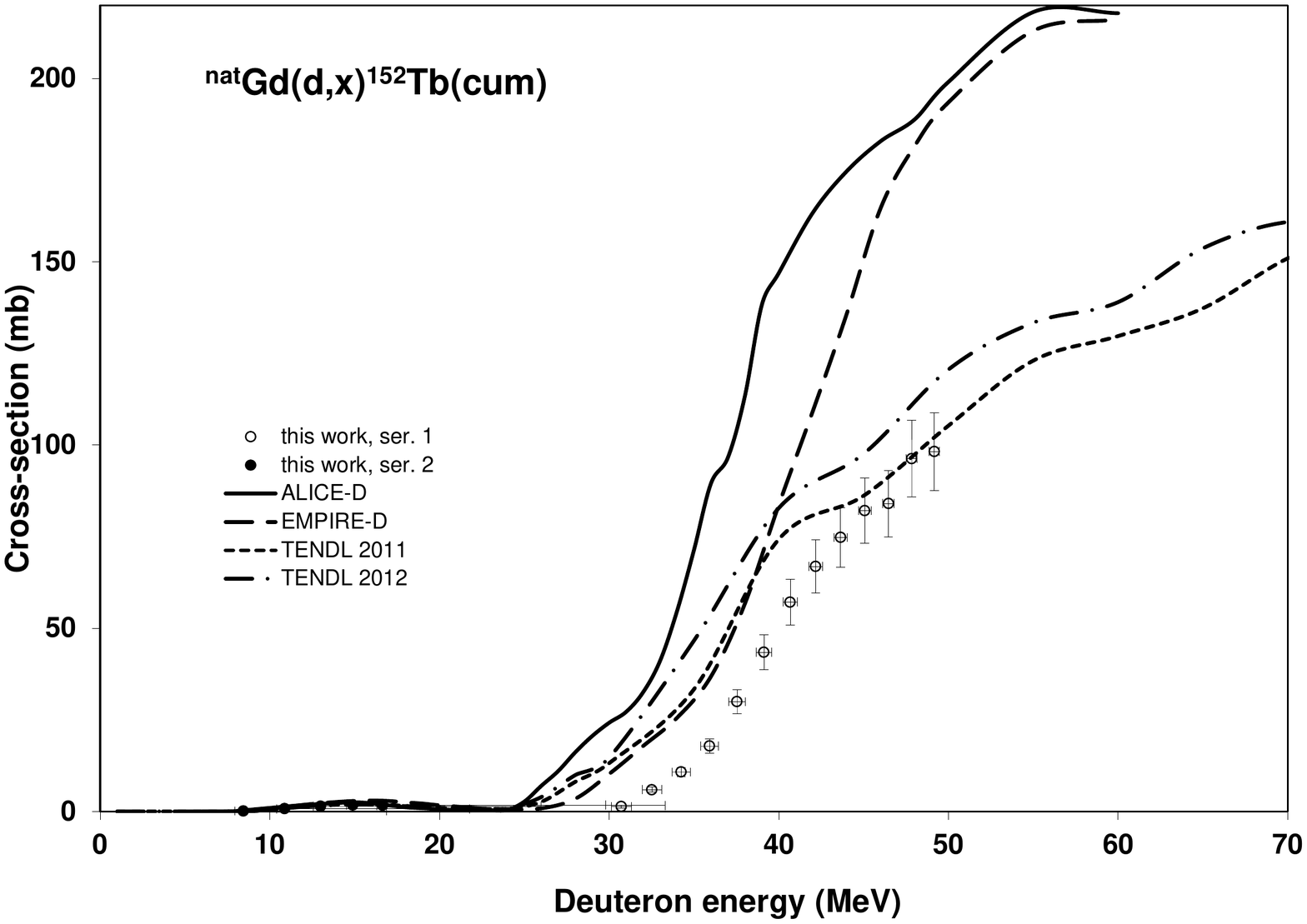}
\caption{Experimental and theoretical excitation function of the $^{nat}Gd$(d,x)$^{152}Tb$ (m+) reaction}
\end{figure}

\subsubsection{Activation cross-sections for production of $^{151}Tb$ (m+) ($T_{1/2}$=17.609 h)}
\label{4.1.10}
The measured excitation function contains the contribution from the decay of the short-lived isomeric state (25 s, IT: 93.4 \%). Above 35 MeV all theoretical predictions overestimate the experimental values (Fig. 10).

\begin{figure}[h]
\includegraphics[scale=0.3]{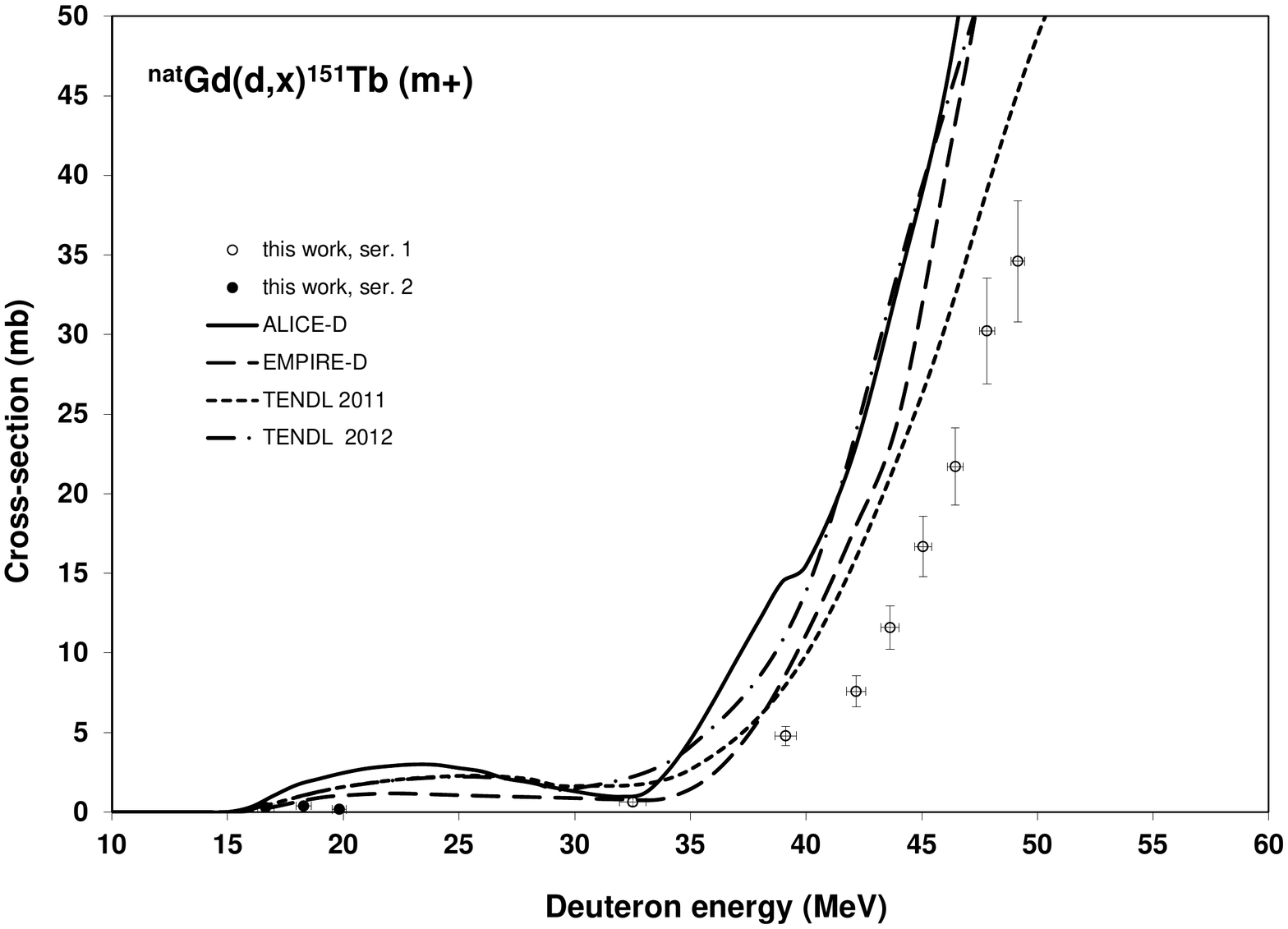}
\caption{Experimental and theoretical excitation function of the $^{nat}Gd$(d,x)$^{151}Tb$ (m+) reaction}
\end{figure}

\subsubsection{Activation cross-sections for production of $^{159}Gd$ ($T_{1/2}$=18.479 h)}
\label{4.1.11}
There are two maxima in the measured excitation function of $^{nat}Gd$(d,x)$^{159}Gd$ (Fig. 11). The first comes from the $^{158}Gd$(d,p)$^{159}Gd$ reaction and the second from nuclear reactions on $^{160}Gd$, namely from $^{160}Gd$(d,p2n)$^{159}Gd$ and from the  $\beta^-$-decay of $^{159}Eu$ (18.1 min), produced via $^{160}Gd$(d,2pn)$^{159}Eu$ reaction. The TENDL data significantly underestimate the $^{158}Gd$(d,p)$^{159}Gd$ reaction, while by ALICE-D and EMPIRE-D (where in the D-versions the (d,p) reaction is enhanced) the higher energy peak is lower than the experiment and the TENDL value.

\begin{figure}[h]
\includegraphics[scale=0.3]{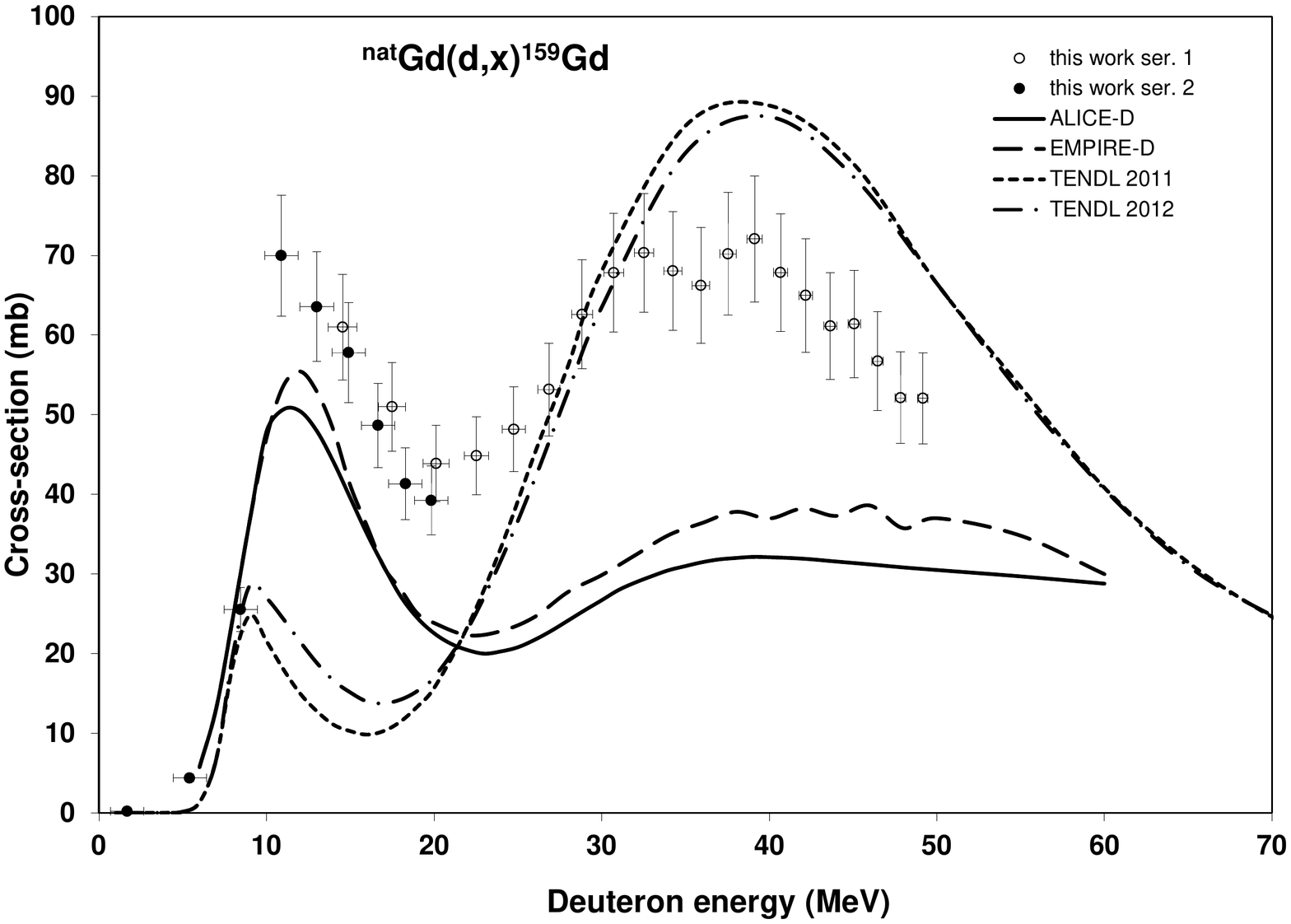}
\caption{Experimental and theoretical excitation function of the $^{nat}Gd$(d,x)$^{159}Gd$ reaction}
\end{figure}

\subsubsection{Activation cross-sections for production of $^{153}Gd$(cum)($T_{1/2}$=240.4 d)}
\label{4.1.12}
The measured excitation function of $^{153}Gd$ (240.4 d) (Fig. 12) was deduced from spectra measured nearly 10 days after EOB, i.e. after five half-lives of the parent $^{153}Tb$ (2.34 d). In these measuring conditions only around 3\% from the decay of $^{153}Tb$ was missing. The missing part was corrected on the basis of the measured cross-section of $^{153}Tb$. The three codes result in acceptable predictions. 

\begin{figure}[h]
\includegraphics[scale=0.3]{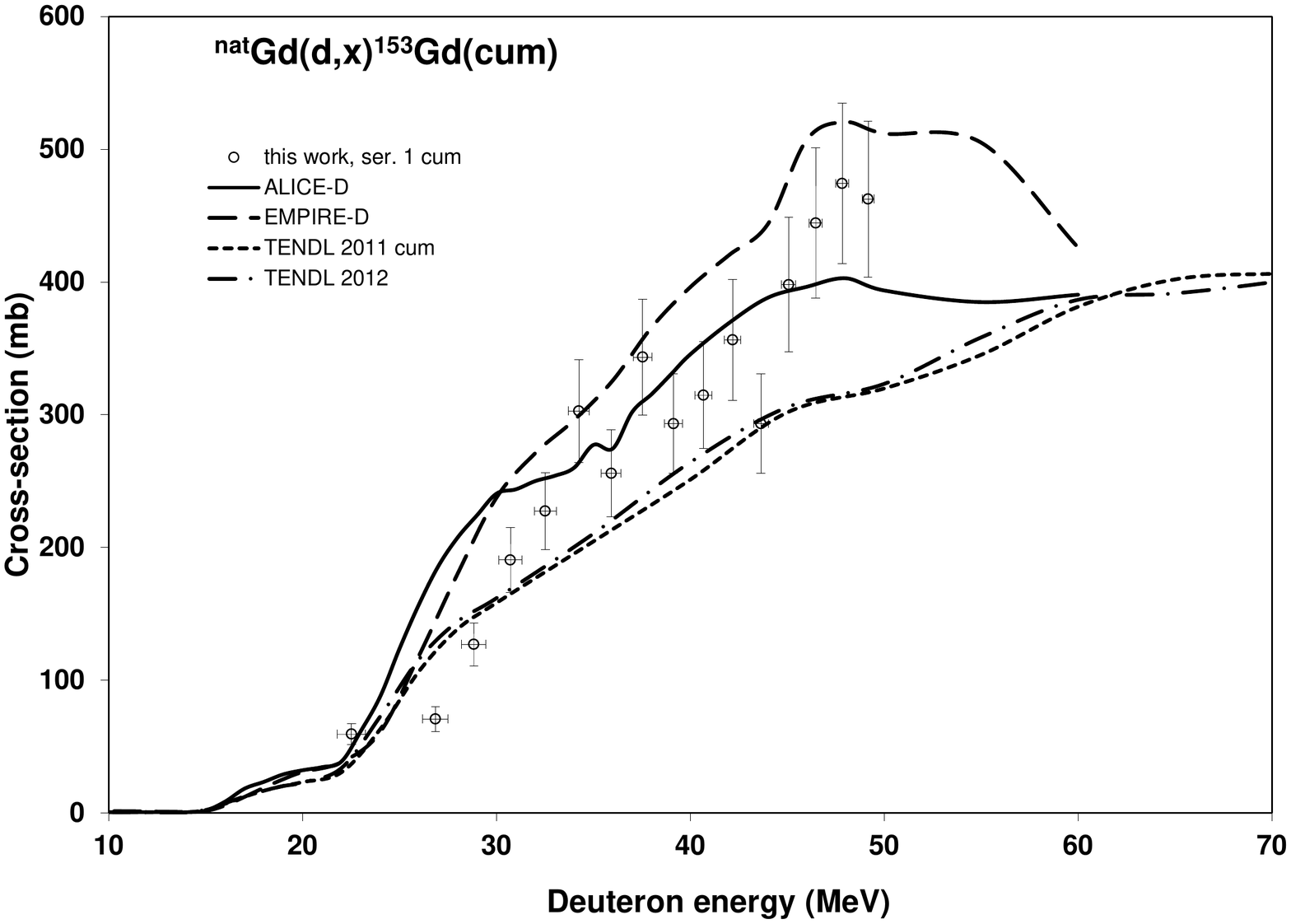}
\caption{Experimental and theoretical excitation function of the $^{nat}Gd$(d,x)$^{153}Gd$(cum) reaction}
\end{figure}

\subsubsection{Activation cross-sections for production of $^{151}Gd$ ($T_{1/2}$=123.9 d)}
\label{4.1.13}
The cumulative cross-sections include the complete decay of $^{151}Tb$ (17.609 h) (Fig. 13). The results of the model calculations generally run together in the investigated energy range, but they more or less overestimate the experimental values.

\begin{figure}[h]
\includegraphics[scale=0.3]{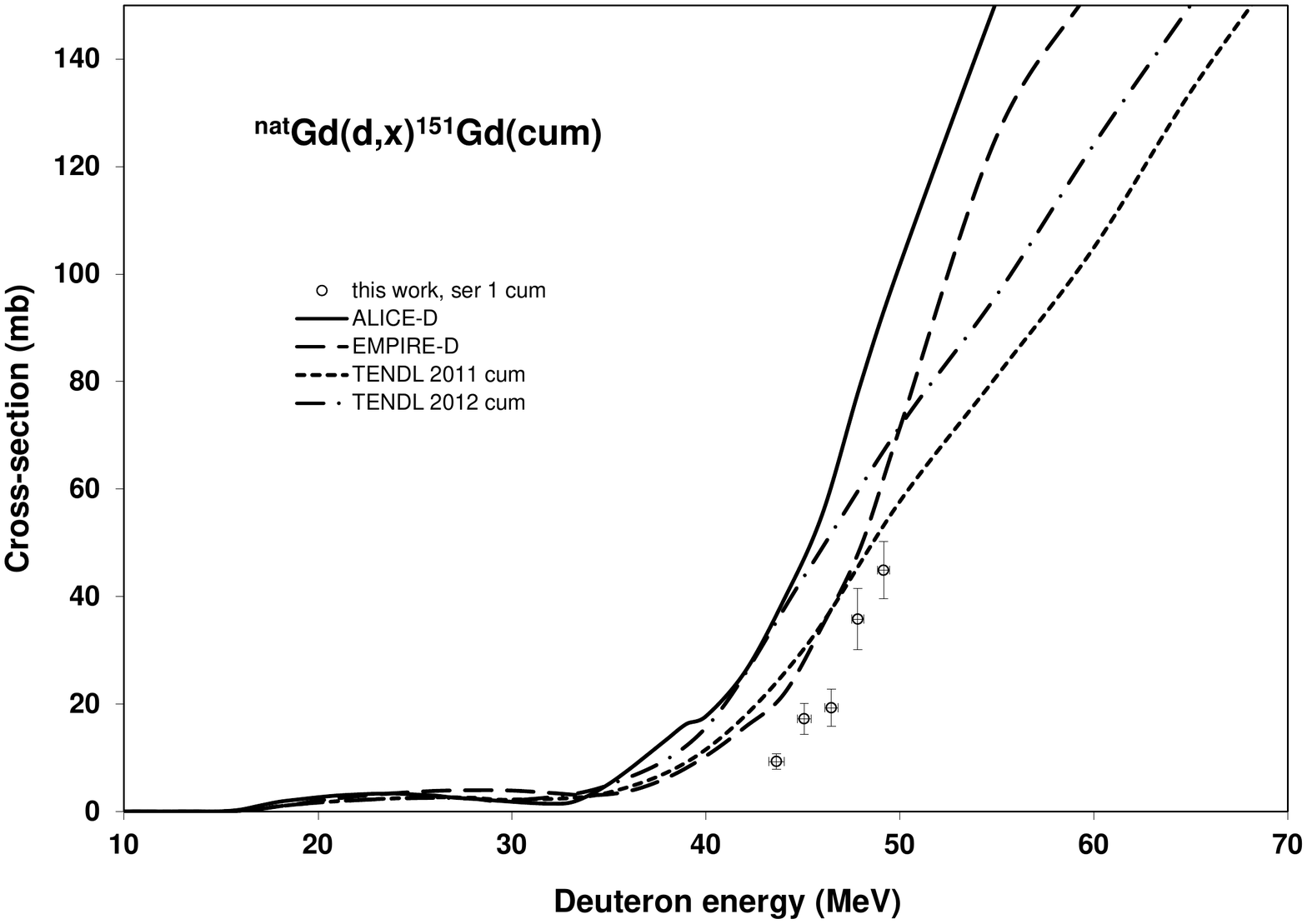}
\caption{Experimental and theoretical excitation function of the $^{nat}Gd$(d,x)$^{151}Gd$ (cum) reaction}
\end{figure}

\subsubsection{Activation cross-sections for production of $^{156}Eu$ ($T_{1/2}$=15.19 d)}
\label{4.1.14}
In the investigated energy range $^{156}Eu$ is produced mostly via (d,$\alpha$xn) reactions. The contribution from the parent 156Sm (9.4 h, $\beta^-$) is small due to the low cross-sections expected for the (d,3pxn) reactions (Fig. 14).

\begin{figure}[h]
\includegraphics[scale=0.3]{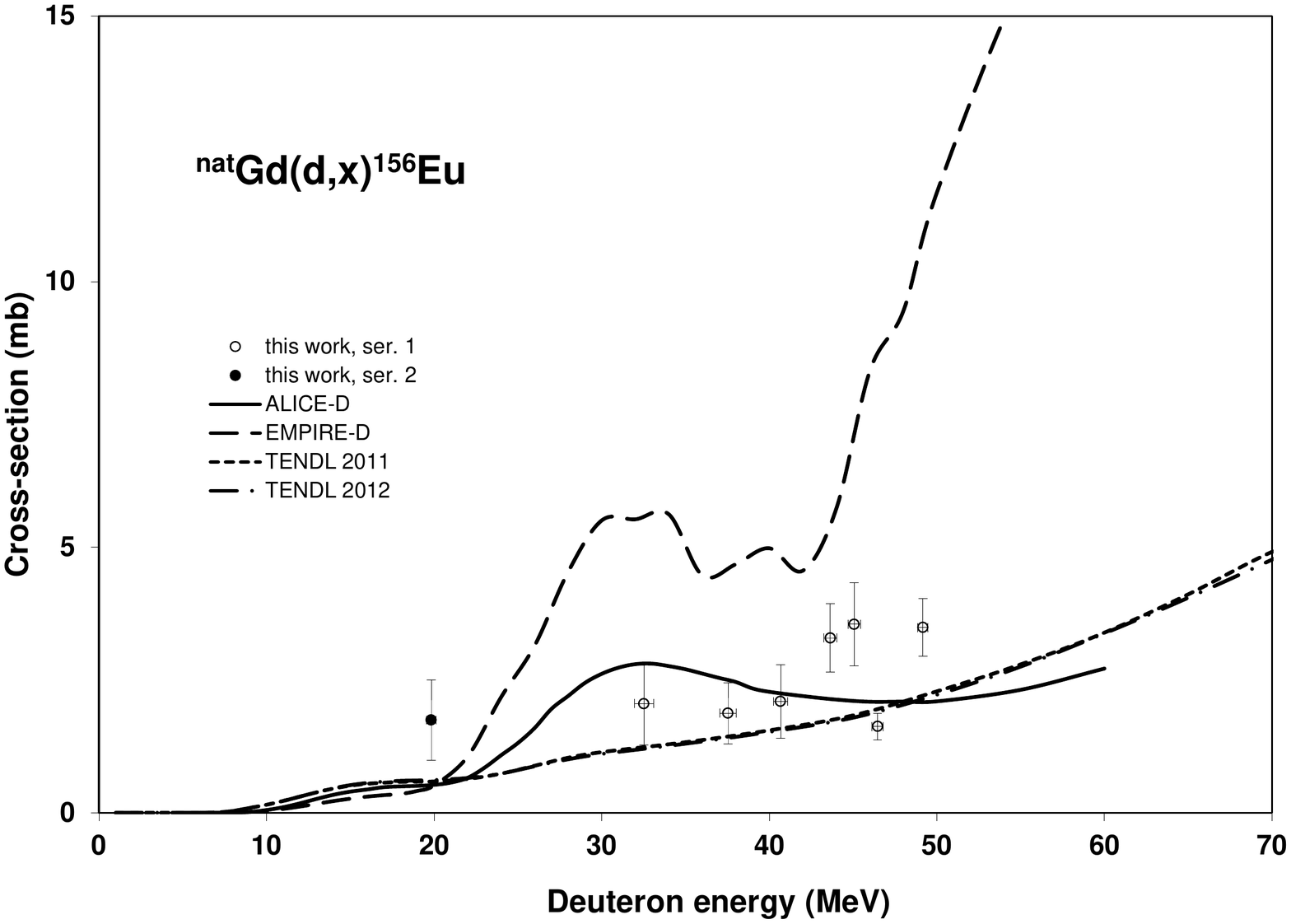}
\caption{Experimental and theoretical excitation function of the $^{nat}Gd$(d,x)$^{156}Eu$ reaction}
\end{figure}

\subsection{Integral yields}
\label{4.2}
From fits to our experimental excitation functions thick target physical yields (instantaneous short irradiation) were calculated. The integral yields are shown in Fig. 15-16 in comparison with the directly measured data \citep{Dmitriev82, Dmitriev89}. The agreement with the existing experimental values for $^{155}Tb$ and $^{156}Tb$ is acceptable (Fig. 15).

\begin{figure}[h]
\includegraphics[scale=0.3]{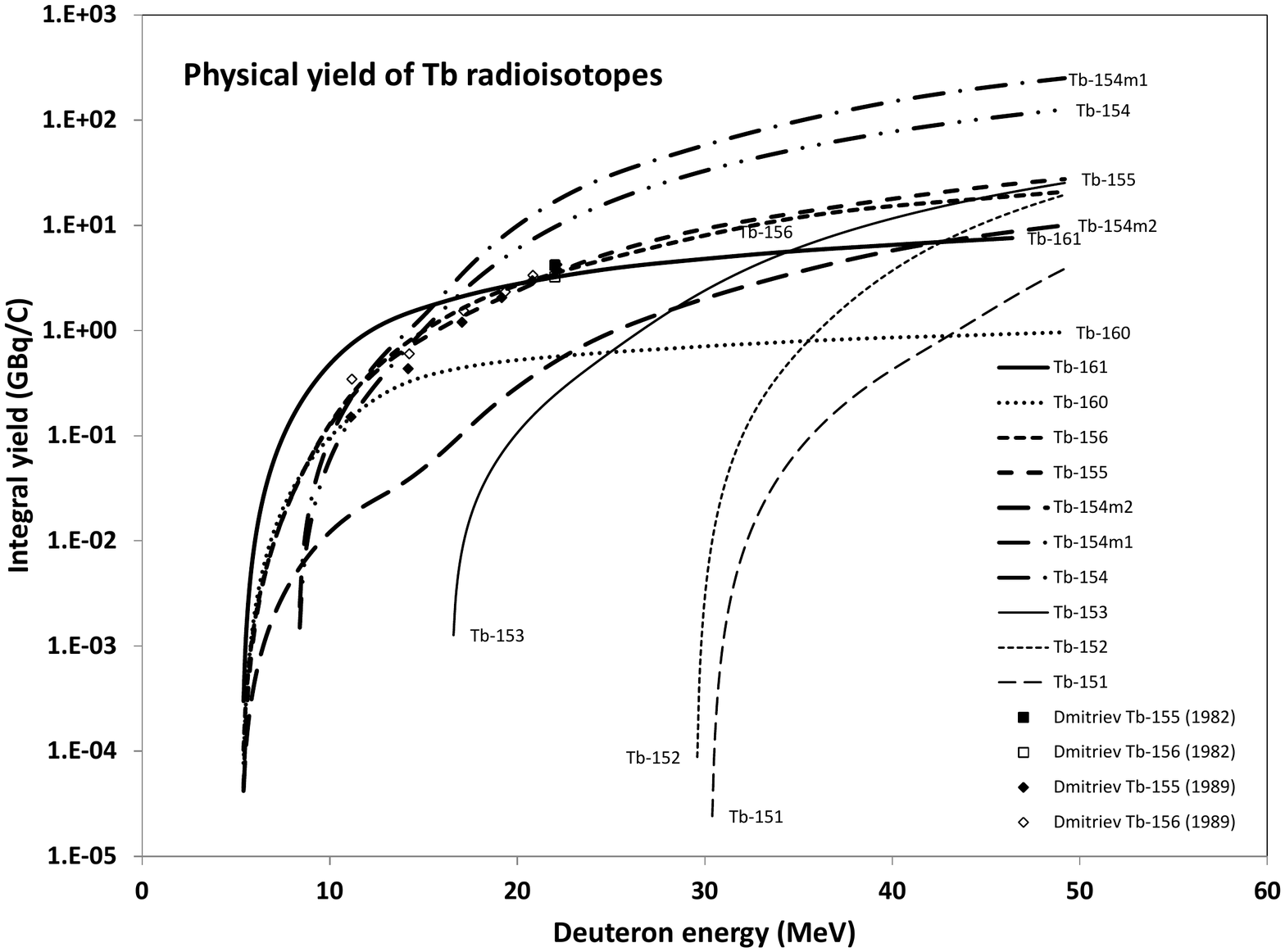}
\caption{Calculated  integral yields of  Tb radioisotopes  on natural gadolinium as a function of the energy}
\end{figure}

\begin{figure}[h]
\includegraphics[scale=0.3]{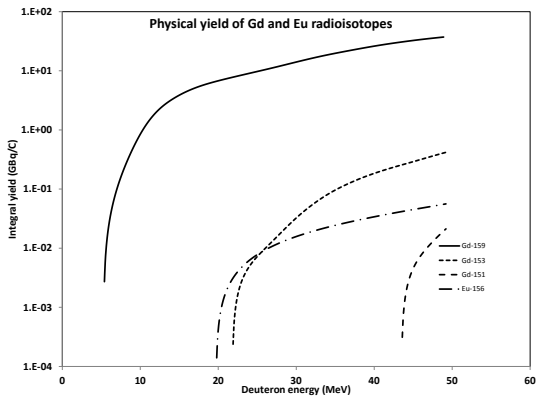}
\caption{Calculated  integral yields of  Gd and Eu radioisotopes on natural gadolinium as a function of the energy}
\end{figure}

\begin{table*}[t]
\tiny
\caption{Experimental cross-sections of ${}^{nat}$Gd(d,xn)${}^{161,160,}$${}^{156}$${}^{(m+),155,}$${}^{154}$${}^{g}$${}^{,154m1,154m2,153,152}$${}^{(m+)}$${}^{,151}$${}^{(m+)}$Tb reactions}
\centering
\begin{center}
\begin{tabular}{|p{0.15in}|p{0.1in}|p{0.2in}|p{0.15in}|p{0.2in}|p{0.15in}|p{0.2in}|p{0.15in}|p{0.2in}|p{0.15in}|p{0.2in}|p{0.15in}|p{0.2in}|p{0.15in}|p{0.2in}|p{0.15in}|p{0.2in}|p{0.15in}|p{0.2in}|p{0.15in}|p{0.2in}|p{0.15in}|} \hline 
\multicolumn{2}{|p{0.25in}|}{E $\pm\Delta$E\newline (MeV)} & \multicolumn{20}{|p{3.5in}|}{Cross-section($\sigma$)$\pm\Delta\sigma$\newline (mb)} \\ \hline 
\multicolumn{2}{|p{0.25in}|}{} & \multicolumn{2}{|p{0.35in}|}{${}^{161}$Tb(cum)} & \multicolumn{2}{|p{0.35in}|}{${}^{160}$Tb} & \multicolumn{2}{|p{0.35in}|}{${}^{156}$Tb(m+)} & \multicolumn{2}{|p{0.35in}|}{${}^{155}$Tb} & \multicolumn{2}{|p{0.35in}|}{${}^{154}$${}^{g}$Tb} & \multicolumn{2}{|p{0.35in}|}{${}^{154m1}$Tb} & \multicolumn{2}{|p{0.35in}|}{${}^{154m2}$Tb} & \multicolumn{2}{|p{0.35in}|}{${}^{153}$Tb} & \multicolumn{2}{|p{0.35in}|}{${}^{152}$Tb(m+)} & \multicolumn{2}{|p{0.35in}|}{${}^{151}$Tb(m+)} \\ \hline 
5.4 & 0.6 & 4.1 & 0.7 & 11.3 & 1.4 & 1.4 & 0.3 & 0.6 & 0.3 & 0.1 & 0.05 &  & ~ &  & ~ &  & ~ &  & ~ &  & ~ \\ \hline 
8.4 & 0.5 & 139 & 16 & 303 & 33 & 33.1 & 4.0 & 32.8 & 4.1 & 0.6 & 0.1 & 1.8 & 0.2 & 2.6 & 0.4 &  & ~ &  & ~ &  & ~ \\ \hline 
10.9 & 0.5 & 234 & 27 & 435 & 48 & 106.6 & 12 & 90.5 & 15.9 & 0.8 & 0.1 & 8.6 & 0.9 & 13.0 & 1.5 &  & ~ &  & ~ &  & ~ \\ \hline 
13.0 & 0.4 & 200 & 23 & 531 & 58 & 144.1 & 16 & 139 & 22 & 0.9 & 0.1 & 15.4 & 1.7 & 23.4 & 2.6 &  & ~ & 0.11 & 0.016 &  & ~ \\ \hline 
13.5 & 0.9 & 223 & 24 & 553 & 61 & 155.7 & 17 & 120 & 13 &  & ~ & 15.1 & 5.2 &  & ~ &  & ~ &  & ~ &  & ~ \\ \hline 
14.9 & 0.4 & 144 & 17 & 346 & 38 & 180.1 & 20 & 140 & 19 & 1.9 & 0.2 & 27.6 & 3.0 & 42.5 & 4.7 &  & ~ & 0.73 & 0.087 &  & ~ \\ \hline 
16.6 & 0.4 & 141 & 18 & 268 & 30 & 246.9 & 27 & 193 & 26 & 3.8 & 0.4 & 61.8 & 6.7 & 102 & 11 & 3.7 & 0.5 & 1.4 & 0.2 & 0.29 & 0.08 \\ \hline 
18.3 & 0.3 & 132 & 18 & 191 & 22 & 252.7 & 28 & 195 & 26 & 6.8 & 0.8 & 93.9 & 10.2 & 138 & 15 & 8.0 & 1.0 & 1.7 & 0.2 & 0.38 & 0.18 \\ \hline 
19.4 & 0.8 & 193 & 21 & 197 & 23 & 244.3 & 26 & 252 & 27 & 7.1 & 0.8 & 110 & 12 & 157 & 17 & 10.9 & 1.2 &  & ~ &  & ~ \\ \hline 
19.8 & 0.3 & 116 & 21 & 144 & 17 & 236.8 & 26 & 211 & 28 & 9.1 & 1.0 & 118 & 13 & 174 & 19 & 13.3 & 1.5 & 1.6 & 0.2 & 0.17 & 0.04 \\ \hline 
21.9 & 0.7 & 160 & 17 & 147 & 20 & 243.2 & 26 & 403 & 44 & 11.5 & 1.3 & 154 & 17 & 207 & 23 & 20.5 & 2.2 &  & ~ &  & ~ \\ \hline 
24.2 & 0.7 & 147 & 16 & 132 & 18 & 244.4 & 26 & 329 & 36 & 14.4 & 1.6 & 169 & 18 & 210 & 23 & 30.7 & 3.4 &  & ~ &  & ~ \\ \hline 
26.4 & 0.7 & 112 & 40 & 106 & 16 & 279.6 & 30 & 365 & 40 & 14.5 & 1.6 & 159 & 17 & 241 & 26 & 53.4 & 5.8 &  & ~ &  & ~ \\ \hline 
28.4 & 0.6 & 104 & 11 & 108 & 17 & 313.1 & 34 & 360 & 39 & 18.3 & 2.0 & 189 & 21 & 271 & 30 & 91.1 & 9.9 &  & ~ &  & ~ \\ \hline 
30.4 & 0.6 & 105 & 11 & 109 & 17 & 330.4 & 36 & 334 & 36 & 19.8 & 2.2 & 225 & 24 & 253 & 28 & 121 & 13 & 1.3 & 0.5 &  & ~ \\ \hline 
32.2 & 0.6 &  & ~ & 80 & 15 & 337.2 & 37 & 324 & 35 &  & ~ &  & ~ & 268 & 35 & 146 & 16 & 5.9 & 0.7 & 0.64 & 0.24 \\ \hline 
34.0 & 0.5 &  & ~ & 100 & 18 & 330.6 & 36 & 328 & 36 & 23.7 & 2.6 & 241 & 26 & 315 & 35 & 152 & 16 & 10.8 & 1.2 &  & ~ \\ \hline 
35.7 & 0.5 & 83.4 & 9 & 74.2 & 19.1 & 298.2 & 32 & 334 & 36 & 26.3 & 2.9 & 261 & 28 & 310 & 34 & 165 & 18 & 17.8 & 2.0 &  & ~ \\ \hline 
37.4 & 0.5 & 72.3 & 29 & 72.4 & 13.6 & 263.7 & 29 & 351 & 38 & 28.6 & 3.1 & 283 & 31 & 319 & 35 & 180 & 20 & 29.9 & 3.3 &  & ~ \\ \hline 
39.0 & 0.5 &  & ~ & 65.8 & 14.2 & 226.9 & 25 & 361 & 39 & 27.1 & 3.0 & 279 & 30 & 313 & 34 & 186 & 20 & 43.4 & 4.7 & 4.8 & 0.6 \\ \hline 
40.5 & 0.4 & 77. & 8.3 & 48.7 & 12.4 & 203.6 & 22 & 366 & 40 & 30.5 & 3.3 & 303 & 33 & 323 & 36 & 215 & 23 & 57.1 & 6.2 &  & ~ \\ \hline 
42.1 & 0.4 &  & ~ & 45.4 & 11.5 & 192.7 & 21 & 376 & 41 & 26.2 & 2.9 & 252 & 27 & 306 & 34 & 204 & 22 & 66.8 & 7.3 & 7.6 & 1.0 \\ \hline 
43.6 & 0.4 &  & ~ & 52.2 & 10.1 & 192.6 & 21 & 372 & 40 & 25.0 & 2.7 & 249 & 27 & 298 & 33 & 214 & 23 & 74.8 & 8.1 & 11.6 & 1.4 \\ \hline 
45.0 & 0.4 & 66.1 & 36.9 & 57.5 & 14.8 & 202.8 & 22 & 367 & 40 & 26.7 & 2.9 & 264 & 29 & 274 & 30 & 232 & 25 & 82.1 & 8.9 & 16.7 & 1.9 \\ \hline 
46.4 & 0.3 & 76.1 & 15.8 & 52.2 & 6.3 & 207.0 & 22 & 332 & 36 & 25.2 & 2.8 & 249 & 27 & 283 & 31 & 227 & 25 & 84.0 & 9.1 & 21.7 & 2.4 \\ \hline 
47.8 & 0.3 &  & ~ & 52.9 & 13.0 & 226.3 & 25 & 320 & 35 & 24.1 & 2.6 & 226 & 24 & 307 & 34 & 213 & 23 & 96.3 & 10.4 & 30.2 & 3.3 \\ \hline 
49.2 & 0.3 &  & ~ & 49.6 & 8.4 & 230.6 & 25 & 285 & 31 & 26.5 & 2.9 & 252 & 27 & 286 & 31 & 232 & 25 & 98.2 & 10.6 & 34.6 & 3.8 \\ \hline 
\end{tabular}

\end{center}
\end{table*}

\begin{table*}[t]
\tiny
\caption{Experimental cross-sections of ${}^{nat}$Gd(d,x)${}^{159,153,151}$Gd and ${}^{nat}$Gd(d,x)${}^{156}$Eu reactions}
\centering
\begin{center}
\begin{tabular}{|p{0.3in}|p{0.3in}|p{0.3in}|p{0.3in}|p{0.3in}|p{0.3in}|p{0.3in}|p{0.3in}|p{0.3in}|p{0.3in}|} \hline 
\multicolumn{2}{|p{1in}|}{E $\pm\Delta$E\newline (MeV)} & \multicolumn{8}{|p{2.5in}|}{Cross-section($\sigma$)$\pm\Delta\sigma$\newline (mb)} \\ \hline 
\multicolumn{2}{|p{1in}|}{} & \multicolumn{2}{|p{0.7in}|}{${}^{159}$Gd} & \multicolumn{2}{|p{0.5in}|}{${}^{153}$Gd} & \multicolumn{2}{|p{0.7in}|}{${}^{151}$Gd} & \multicolumn{2}{|p{0.7in}|}{${}^{156}$Eu} \\ \hline 
5.4 & 0.6 & 4.37 & 0.49 &  & ~ &  & ~ &  & ~ \\ \hline 
8.4 & 0.5 & 25.5 & 2.8 &  & ~ &  & ~ &  & ~ \\ \hline 
10.9 & 0.5 & 70.0 & 7.6 &  & ~ &  & ~ &  & ~ \\ \hline 
13.0 & 0.4 & 63.5 & 6.9 &  & ~ &  & ~ &  & ~ \\ \hline 
13.5 & 0.9 & 61.0 & 6.6 &  & ~ &  & ~ &  & ~ \\ \hline 
14.9 & 0.4 & 57.8 & 6.3 &  & ~ &  & ~ &  & ~ \\ \hline 
16.6 & 0.4 & 48.7 & 5.3 &  & ~ &  & ~ &  & ~ \\ \hline 
18.3 & 0.3 & 41.3 & 4.5 &  & ~ &  & ~ &  & ~ \\ \hline 
19.4 & 0.8 & 43.9 & 4.8 &  & ~ &  & ~ &  & ~ \\ \hline 
19.8 & 0.3 & 39.2 & 4.3 &  & ~ &  & ~ & 1.74 & 0.76 \\ \hline 
21.9 & 0.7 & 44.8 & 4.9 & 59.3 & 7.8 &  & ~ &  & ~ \\ \hline 
24.2 & 0.7 & 48.2 & 5.3 &  & ~ &  & ~ &  & ~ \\ \hline 
26.4 & 0.7 & 53.2 & 5.8 & 70.8 & 9.3 &  & ~ &  & ~ \\ \hline 
28.4 & 0.6 & 62.6 & 6.8 & 127 & 16 &  & ~ &  & ~ \\ \hline 
30.4 & 0.6 & 67.8 & 7.4 & 191 & 24 &  & ~ &  & ~ \\ \hline 
32.2 & 0.6 & 70.4 & 7.5 & 227 & 29 &  & ~ & 2.05 & 0.78 \\ \hline 
34.0 & 0.5 & 68.1 & 7.5 & 303 & 39 &  & ~ &  & ~ \\ \hline 
35.7 & 0.5 & 66.2 & 7.3 & 256 & 33 &  & ~ &  & ~ \\ \hline 
37.4 & 0.5 & 70.2 & 7.7 & 343 & 44 &  & ~ & 1.87 & 0.58 \\ \hline 
39.0 & 0.5 & 72.1 & 7.9 & 293 & 37 &  & ~ &  & ~ \\ \hline 
40.5 & 0.4 & 67.8 & 7.4 & 315 & 40 &  & ~ & 2.09 & 0.69 \\ \hline 
42.1 & 0.4 & 65.0 & 7.1 & 356 & 45 &  & ~ &  & ~ \\ \hline 
43.6 & 0.4 & 61.1 & 6.7 & 293 & 37 & 9.3 & 1.5 & 3.29 & 0.65 \\ \hline 
45.0 & 0.4 & 61.4 & 6.8 & 398 & 51 & 17.2 & 2.8 & 3.54 & 0.78 \\ \hline 
46.4 & 0.3 & 56.7 & 6.2 & 444 & 57 & 19.3 & 3.5 & 1.62 & 0.25 \\ \hline 
47.8 & 0.3 & 52.1 & 5.7 & 474 & 60 & 35.8 & 5.7 &  & ~ \\ \hline 
49.2 & 0.3 & 52.1 & 5.7 & 463 & 59 & 44.9 & 5.3 & 3.49 & 0.54 \\ \hline 
\end{tabular}

\end{center}
\end{table*}

\begin{table*}[t]
\tiny
\caption{Main light ion induced reactions for direct and indirect production of the ${}^{153}$Gd}
\centering
\begin{center}
\begin{tabular}{|p{1.2in}|p{0.8in}|p{1.0in}|p{1.0in}|p{0.6in}|} \hline 
Reaction & Energy range (MeV) & Yield\newline (MBq/C) & Disturbing reaction & Threshold\newline (MeV) \\ \hline 
${}^{153}$Eu(p,n)${}^{153}$Gd   & 16-6 & 52 & ${}^{153}$Eu(p,3n)${}^{151}$Gd & 16.2 \\ \hline 
${}^{153}$Eu(d,2n)${}^{153}$Gd & 19-7 & 188 & ${}^{153}$Eu(d,4n)${}^{151}$Gd & 18.6 \\ \hline 
${}^{154}$Gd(p,2n)${}^{153}$Tb$\longrightarrow{}^{153}$Gd & 27-12 & 651 & ${}^{154}$Gd(p,4n)${}^{151}$Tb & 27.3~ \\ \hline 
${}^{154}$Gd(d,3n)${}^{153}$Tb$\longrightarrow{}^{153}$Gd & 30-15 & 567 & ${}^{154}$Gd(d,5n)${}^{151}$Tb & 29.7 \\ \hline 
${}^{151}$Eu($\alpha$,2n)${}^{153}$Tb$\longrightarrow{}^{153}$Gd & 34-19 & 31 & ${}^{151}$Eu($\alpha$,4n)${}^{151}$Tb & 33.8~ \\ \hline 
${}^{151}$Eu(${}^{3}$He,n)${}^{153}$Tb$\longrightarrow{}^{153}$Gd & 13-0 & 10${}^{-4}$ & ${}^{151}$Eu(${}^{3}$He,3n)${}^{151}$Tb & 12.6 \\ \hline 
\end{tabular}
\end{center}
\end{table*}

\section{Comparison of the production routes of the $^{153}Gd$}
\label{5}
$^{153}Gd$ (240.4 d) emits two low-energy photons with energies of 97.43 keV and 103.18 keV respectively, which are optimal to penetrate through the body, expose the patient only to limited dose and can be detected with standard imaging technology. 
The $^{153}Gd$ is used in nuclear medicine for attenuation corrections by means of additional  $\gamma$-transmission measurements \citep{Vanlaere}, for determination of the bone mineral content \citep{Smith}, for labeling gadolinium-based contrast agent \citep{Wadas} as well as for localizing sentinel lymph nodes \citep{Liu}.
The long-lived radionuclide $^{153}Gd$ can be produced via various nuclear reactions. Taking into account the long half-life, the traditional way is using parallel production by neutron induced reactions at research reactors. A few neutron induced production routes were previously investigated and reported in the literature:
\begin{itemize}
\item	$^{151}Eu$(n,$\gamma$)$^{152}Eu\longrightarrow ^{152}Gd(n,\gamma)^{153}Gd$ \citep{Karelin }
\item	$^{152}Gd$(n,$\gamma$) on highly enriched $^{152}Gd$  \citep{Case}
\item	$^{156}Dy$(n,$\alpha$)$^{153}Gd$ reaction \citep{Waheed }
\end{itemize}
The first two methods result in carrier added product, but by using high flux reactors and long irradiations the specific activity can be rather high. The (n,$\alpha$) reaction (Q = 8.006 MeV) results in no-carrier added product, but the yield is also low.
There are various production routes using charged particle accelerators for production of no carrier added $^{153}Gd$:
\begin{itemize}
\item	Spallation reactions using mass separators \citep{Beyer}
\item	Low energy light ion induced reactions
\end{itemize}
By using light ion induced reactions the $^{153}Gd$ (240.4 d) can be produced directly or through the decay of the $^{153}Tb$ (2.34 d). The most important direct and indirect reactions, together with the possible disturbing reactions are collected in Table 6. 
For the $^{153}Eu$(p,n)$^{153}Gd$  and $^{153}Eu$(d,2n)$^{153}Gd$ nuclear reactions experimental excitation functions are available \citep{West}. For other reactions the TALYS data from the TENDL-2012 library are presented. For comparison, the predictive force of the TENDL data is satisfactory (see also Fig. 12), taking into account the resulted large differences in the production yields of the different production routes. The $^{153}Eu$(p,n) and $^{153}Eu$(d,2n) reactions result in direct production. After irradiation the Gd can be separated from the target. In the case of $^{154}Gd$(p,2n)$^{153}Tb\longrightarrow^{153}Gd$, $^{154}Gd$(d,3n)$^{153}Tb\longrightarrow ^{153}Gd$, $^{151}Eu$($\alpha$ ,2n)$^{153}Tb\longrightarrow ^{153}Gd$  and $^{151}Eu(^3He$,n)$^{153}Tb\longrightarrow^{153}Gd$  routes the production  is indirect, the produced $^{153}Tb$ must be separated at the end of the irradiation from the Gd or Eu target to assure  no carrier added final product. The excitation functions are shown on Fig. 17.
In all cases, from the point of view of radionuclidic purity, the production of the $^{151}Gd$ (120 d) and consequently also its parent $^{151}Tb$ (17.609 h) should be eliminated, which gives an upper limit for the energy of the bombarding beam (see the energy windows in Table 6).
The comparison of the productivity for different charged particle induced reaction shows that in the case of direct production the deuterons are more favorable (a factor of 3, see Table 6). The energy windows were chosen in such a way that the production and target preparation is optimal, based on the cross-sections, thresholds of the disturbing reactions and also on experimental experiences. In the case of indirect production however, the protons are slightly more favorable. The yield of the alpha-particle induced reaction is significantly less than that of the deuteron induced reactions. The yield of the $^3He$ route is negligible ( max=1.5 mb, Y = 104 MBq/C). The indirect proton and deuteron yields are comparable, the proton irradiation yields 20\% more $^{153}Gd$, and it must also be mentioned that less accelerators can provide 30 MeV deuteron energy.	
There are many other factors determining the final competitivity of the different production routes:
\begin{itemize}
\item	Taking into account the long half-life of $^{153}Gd$, long irradiations are necessary to produce reasonable activities, which is straightforward in the case of direct reactions. Using the indirect route the half-life of the $^{153}Tb$ limits the length of a single irradiation. Moreover, the repeated separation causes losses in the expensive enriched target material
\item	The price of the highly enriched $^{153}Eu$ is significantly lower compared to that of highly enriched $^{154}Gd$ 
\item	In general, cyclotrons have significantly lower beam intensities for d-, $\alpha$ - and $^3He$-particle beams compared to protons
\end{itemize}

There are other important factors, when one compares with the production routes at research reactors.

\begin{itemize}
\item	The long half-life requires long irradiations. The most economical way of it is simultaneous parasitic irradiation (dividing the beam), which in case of low and medium energy cyclotrons is not a simple task (cooling problems, target construction, etc.), but in the case of reactors it is an everyday practice
\item	The cross-sections of the (n,$\gamma$) reactions are significantly higher compared to the charged particle induced reactions
\item	In case of neutrons large mass targets can be used, without beam stopping and cooling problems
\end{itemize}

By summarizing the above mentioned factors, it can be concluded that it is very difficult to compete with the reactor production. The charged particle routes can only be used for production of small amounts for research purposes.	

\begin{figure}[h]
\includegraphics[scale=0.3]{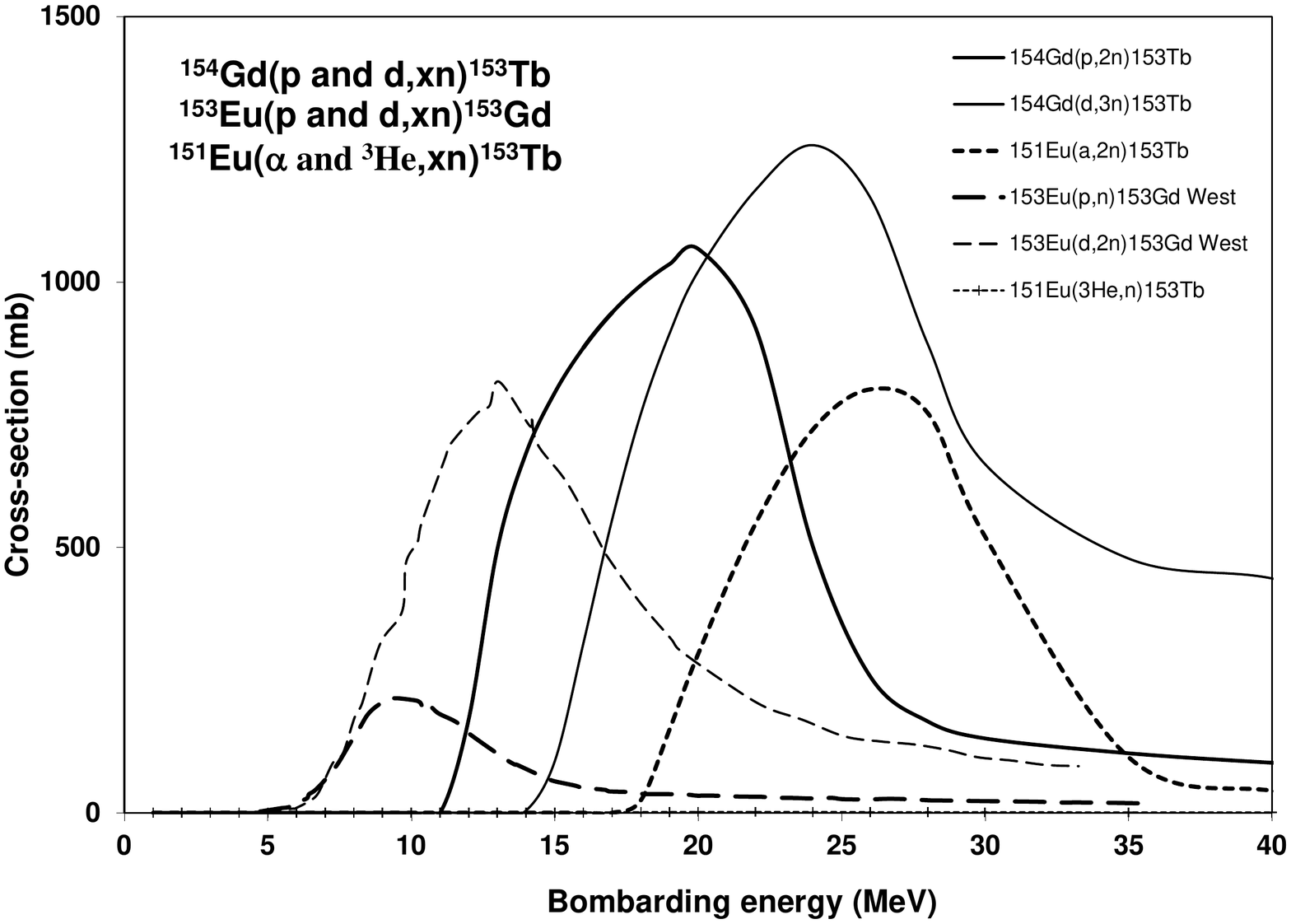}
\caption{Comparison of the excitation functions of the light ion induced production routes of $^{153}Gd$ and $^{153}Tb$ by using literature data or model calculations}
\end{figure}

\section{Summary and conclusions}
\label{6}
This work was performed in the frame of the systematic study of activation cross-sections for applications and for development of the nuclear reaction model codes. 
In this study experimental cross-sections for the $^{nat}Gd$(d,xn)$^{161,160,156(m+),154,154m1,154m2,153,152(m+),151(m+)}Tb$, $^{nat}Gd$(d,xn)$^{159,153,151}Gd$ and $^{nat}Gd$(d,x)$^{156}Eu$ nuclear  reactions were measured up to 50 MeV. Comparison of the experimental and theoretical results calculated by the ALICE-D EMPIRE-D and TALYS codes shows that still significant disagreements between the theoretical predictions and the existing experimental data. The empirical improvement of the deuteron breakup description, which was a serious problem by the original EMPIRE and ALICE codes by describing deuteron induced reactions, has improved the predictions in the EMPIRE-D and ALICE-D codes. 
Among the studied reaction products $^{161}Tb$ and $^{153}Gd$ are of importance for nuclear medicine. 
In order to prepare a carrier free $^{161}Tb$ end product the $^{160}Gd$(n, )$^{161}Gd$?$^{161}Tb$ route is more productive than $^{160}Gd$(d,n)$^{161}Tb$+ $^{160}Gd$(d,p)$^{161}Gd\longrightarrow^{161}Tb$ \citep{TF2013}.
According to our comparison of production routes of carrier free $^{153}Gd$ (this work) the deuteron induced reactions on Eu and Gd are competitive with proton induced reactions in special cases concerning to yields, but because of the availability of the required deuteron energy and intensity, the proton induced reaction is more favorable.

\section{Acknowledgements}
\label{7}
This work was performed in the frame of the HAS-FWO Vlaanderen (Hungary-Belgium) project. The authors acknowledge the support of the research project and of the respective institutions. We are grateful to all the authorities concerned. We thank to the Cyclotron Laboratory of the Université Catholique in Louvain la Neuve (LLN) providing the beam time and the staff of the LLN Cyclone 90 cyclotron for performing the irradiations.
 


%
%

\clearpage
\bibliographystyle{elsarticle-harv}
\bibliography{Gdd}

\begin{thebibliography}{30}
\expandafter\ifx\csname natexlab\endcsname\relax\def\natexlab#1{#1}\fi
\expandafter\ifx\csname url\endcsname\relax
  \def\url#1{\texttt{#1}}\fi
\expandafter\ifx\csname urlprefix\endcsname\relax\def\urlprefix{URL }\fi

\bibitem[{Andersen and Ziegler(1977)}]{Andersen}
Andersen, H.~H., Ziegler, J.~F., 1977. Hydrogen stopping powers and ranges in
  all elements. The Stopping and ranges of ions in matter, Volume 3. The
  Stopping and ranges of ions in matter. Pergamon Press, New York.

\bibitem[{Beyer and Ruth(2003)}]{Beyer}
Beyer, G.~J., Ruth, T.~J., 2003. The role of electromagnetic separators in the
  production of radiotracers for bio-medical research and nuclear medical
  application. Nuclear Instruments \& Methods in Physics Research Section
  B-Beam Interactions with Materials and Atoms 204, 694--700.

\bibitem[{Bonardi(1987)}]{Bonardi}
Bonardi, M., 1987. The contribution to nuclear data for biomedical radioisotope
  production from the milan cyclotron facility.

\bibitem[{Canberra(2000)}]{Canberra}
Canberra, 2000.
  http://www.canberra.com/products/radio-chemistry\_lab/genie-2000-software.asp.

\bibitem[{Case et~al.(1969)Case, Acree, and Cutshal}]{Case}
Case, F.~N., Acree, E.~H., Cutshal, N.~H., 1969. Production study of
  gadolinium-153. Tech. rep.

\bibitem[{Dityuk et~al.(1998)Dityuk, Konobeyev, Lunev, and Shubin}]{Dityuk}
Dityuk, A.~I., Konobeyev, A.~Y., Lunev, V.~P., Shubin, Y.~N., 1998. New version
  of the advanced computer code alice-ippe. Tech. rep., IAEA.

\bibitem[{Dmitriev et~al.(1982)Dmitriev, Krasnov, and Molin}]{Dmitriev82}
Dmitriev, P.~P., Krasnov, N.~N., Molin, G.~A., 1982. Radioactive nuclide yields
  for thick target at 22 mev deuterons energy. Yadernie Konstanti 34~(4), 38.

\bibitem[{Dmitriev et~al.(1989)Dmitriev, Molin, and Dmitrieva}]{Dmitriev89}
Dmitriev, P.~P., Molin, G.~A., Dmitrieva, Z.~P., 1989. Production of tb-155 for
  nuclear-medicine in the reactions gd-155(pn), gd-156(p2n), and gd-155(d2n).
  Soviet Atomic Energy 66~(6), 470--472.

\bibitem[{Ekström and Firestone(1999)}]{Ekstrom}
Ekström, L.~P., Firestone, R.~B., 1999. Www table of radioactive isotopes,
  database version 2/28/99.

\bibitem[{Herman et~al.(2007)Herman, Capote, Carlson, Oblozinsky, Sin, Trkov,
  Wienke, and Zerkin}]{Herman}
Herman, M., Capote, R., Carlson, B.~V., Oblozinsky, P., Sin, M., Trkov, A.,
  Wienke, H., Zerkin, V., 2007. Empire: Nuclear reaction model code system for
  data evaluation. Nuclear Data Sheets 108~(12), 2655--2715.

\bibitem[{Hermanne et~al.(2012)Hermanne, Rebeles, T\'ark\'anyi, Tak\'acs,
  Tak\'acs, Csikai, and Ignatyuk}]{Hermanne2012}
Hermanne, A., Rebeles, R.~A., T\'ark\'anyi, F., Tak\'acs, S., Tak\'acs, M.~P.,
  Csikai, J., Ignatyuk, A., 2012. Cross sections of deuteron induced reactions
  on sc-45 up to 50 mev: Experiments and comparison with theoretical codes.
  Nuclear Instruments \& Methods in Physics Research Section B-Beam
  Interactions with Materials and Atoms 270, 106--115.

\bibitem[{Ignatyuk(2011)}]{Ignatyuk}
Ignatyuk, A.~V., 2011. Phenomenological systematics of the (d,p) cross
  sections,
  http://www-nds.iaea.org/fendl3/000pages/rcm3/slides//ignatyuk\_fendl-3

\bibitem[{{International-Bureau-of-Weights-and-Measures}(1993)}]{Error}
{International-Bureau-of-Weights-and-Measures}, 1993. Guide to the expression
  of uncertainty in measurement, 1st Edition. International Organization for
  Standardization, Genève, Switzerland.

\bibitem[{Kinsey et~al.(1997)Kinsey, Dunford, Tuli, and Burrows}]{Kinsey}
Kinsey, R.~R., Dunford, C.~L., Tuli, J.~K., Burrows, T.~W., 1997. Nudat 2.6.
  In: Proceedings of the 9th International Symposium on Capture Gamma – Ray
  Spectroscopy and Related Topics. Vol.~2. Springer Hungarica Ltd, p. 657.

\bibitem[{Koning and Rochman(2012)}]{Koning}
Koning, A.~J., Rochman, D., 2012. Modern nuclear data evaluation with the talys
  code system. Nuclear Data Sheets 113, 2841.

\bibitem[{Koning et~al.(2012)Koning, Rochman, van~der Marck, Kopecky, Sublet,
  Pomp, Sjostrand, Forrest, Bauge, and Henriksson}]{KoningTALYS}
Koning, A.~J., Rochman, D., van~der Marck, S., Kopecky, J., Sublet, J.~C.,
  Pomp, S., Sjostrand, H., Forrest, R., Bauge, E., Henriksson, H., 2012.
  Tendl-2012: Talys-based evaluated nuclear data library.

\bibitem[{Liu et~al.(2013)Liu, Solomon, and Achilefu}]{Liu}
Liu, Y., Solomon, M., Achilefu, S., 2013. Perspectives and potential
  applications of nanomedicine in breast and prostate cancer. Medicinal
  Research Reviews 33~(1), 3--32.

\bibitem[{M\"uller et~al.(2012)M\"uller, Zhernosekov, Köster, Johnston,
  Dorrer, Hohn, van~der Walt, Türler, and Schibli}]{Muller}
M\"uller, C., Zhernosekov, K., Köster, U., Johnston, K., Dorrer, H., Hohn, A.,
  van~der Walt, N.~T., Türler, A., Schibli, R., 2012. A unique matched
  quadruplet of terbium radioisotopes for pet and spect and for α- and
  β−-radionuclide therapy: An in vivo proof-of-concept study with a new
  receptor-targeted folate derivative. The Journal of Nuclear Medicine 53~(12),
  1951--1959.

\bibitem[{NuDat(2011)}]{NuDat}
NuDat, 2011. Nudat2 database http://www.nndc.bnl.gov/nudat2/.

\bibitem[{Pritychenko and Sonzogni(2003)}]{Pritychenko}
Pritychenko, B., Sonzogni, A., 2003. Q-value calculator.

\bibitem[{Rastrepina(2009)}]{Rastrepina}
Rastrepina, G., 2009. Half –life measurements of the $^{154}tb$ isotope.
  Uzhhorod University Scientific Herald. Series Physics. 24, 171--174.

\bibitem[{Smith et~al.(1983)Smith, Sutton, and Tothill}]{Smith}
Smith, M.~A., Sutton, D., Tothill, P., 1983. Comparison between gd-153 and
  am-241, cs-137 for dual-photon absorptiometry of the spine. Physics in
  Medicine and Biology 28~(6), 709--721.

\bibitem[{Sz\'ekely(1985)}]{Szekely}
Sz\'ekely, G., 1985. Fgm - a flexible gamma-spectrum analysis program for a
  small computer. Computer Physics Communications 34~(3), 313--324.

\bibitem[{T\'ark\'anyi et~al.(2007)T\'ark\'anyi, Hermanne, Kir\'aly, Tak\'acs,
  Ditr\'oi, Baba, Ohtsuki, Kovalev, and Ignatyuk}]{TF2007}
T\'ark\'anyi, F., Hermanne, A., Kir\'aly, B., Tak\'acs, S., Ditr\'oi, F., Baba,
  M., Ohtsuki, T., Kovalev, S.~F., Ignatyuk, A.~V., 2007. Study of activation
  cross-sections of deuteron induced reactions on erbium: Production of
  radioisotopes for practical applications. Nuclear Instruments \& Methods in
  Physics Research Section B-Beam Interactions with Materials and Atoms
  259~(2), 829--835.

\bibitem[{T\'ark\'anyi et~al.(2013)T\'ark\'anyi, Hermanne, Tak\'acs, Ditr\'oi,
  Csikai, and Ignatyuk}]{TF2013}
T\'ark\'anyi, F., Hermanne, A., Tak\'acs, S., Ditr\'oi, F., Csikai, J.,
  Ignatyuk, A.~V., 2013. Cross-section measurement of some deuteron induced
  reactions on 160gd for possible production of the therapeutic radionuclide
  161tb. Journal of Radioanalytical and Nuclear Chemistry DOI
  10.1007/s10967-013-2507-x.

\bibitem[{T\'ark\'anyi et~al.(1991)T\'ark\'anyi, Szelecs\'enyi, and
  Tak\'acs}]{TF1991}
T\'ark\'anyi, F., Szelecs\'enyi, F., Tak\'acs, S., 1991. Determination of
  effective bombarding energies and fluxes using improved stacked-foil
  technique. Acta Radiologica, Supplementum 376, 72.

\bibitem[{T\'ark\'anyi et~al.(2001)T\'ark\'anyi, Tak\'acs, Gul, Hermanne,
  Mustafa, Nortier, Oblozinsky, Qaim, Scholten, Shubin, and Youxiang}]{TF2001}
T\'ark\'anyi, F., Tak\'acs, S., Gul, K., Hermanne, A., Mustafa, M.~G., Nortier,
  M., Oblozinsky, P., Qaim, S.~M., Scholten, B., Shubin, Y.~N., Youxiang, Z.,
  2001. Beam monitor reactions (chapter 4). charged particle cross-section
  database for medical radioisotope production: diagnostic radioisotopes and
  monitor reactions. Tech. rep., IAEA.

\bibitem[{Van~Laere et~al.(2000)Van~Laere, Koole, Kauppinen, Monsieurs,
  Bouwens, and Dierck}]{Vanlaere}
Van~Laere, K., Koole, M., Kauppinen, T., Monsieurs, M., Bouwens, L., Dierck,
  R., 2000. Nonuniform transmission in brain spect using 201tl, 153gd, and
  99mtc static line sources: anthropomorphic dosimetry studies and influence on
  brain quantification. Journal of Nuclear Medicine 41~(12), 2051--62.

\bibitem[{Wadas et~al.(2010)Wadas, Sherman, Miner, Duncan, and
  Anderson}]{Wadas}
Wadas, T.~J., Sherman, C.~D., Miner, J.~H., Duncan, J.~R., Anderson, C.~J.,
  2010. The biodistribution of [gd-153]gd-labeled magnetic resonance contrast
  agents in a transgenic mouse model of renal failure differs greatly from
  control mice. Magnetic Resonance in Medicine 64~(5), 1274--1280.

\bibitem[{West et~al.(1993)West, Lanier, Mustafa, Nuckolls, Frehaut, Adam, and
  Philis}]{West}
West, H.~I., Lanier, R.~G., Mustafa, M.~G., Nuckolls, R.~N., Frehaut, J., Adam,
  A., Philis, C.~A., 1993. Proton and deutron excitation functions for eu-151.
  Tech. rep.

\end{thebibliography}




%



\end{document}